\documentclass[12pt]{iopart}
\usepackage{iopams}
\usepackage{graphicx}
\usepackage{soul}
\usepackage[sort,compress]{cite}
\usepackage{siunitx}
\begin{document}

\title[Demystifying the measurement of periodic structures]{Demystifying the measurement of periodic structures}

\iftrue
\author{D Ne\v{c}as$^{1}$,
A Yacoot$^2$
and
P Klapetek$^{1,3}$
}
\address{$^1$ CEITEC, Brno University of Technology, Purky\v{n}ova 123, 61200 Brno, Czech Republic}
\address{$^2$ National Physical Laboratory, Hampton Road, Teddington, Middlesex TW11 0LW United Kingdom.}
\address{$^3$ Czech Metrology Institute, Okru\v{z}n\'{\i} 31, 63800 Brno, Czech Republic}
\eads{\mailto{yeti@physics.muni.cz}, \mailto{pklapetek@cmi.cz}, \mailto{andrew.yacoot@npl.co.uk}}
\fi

\begin{abstract}
Periodic structures are often found in various areas of nanoscience and nanotechnology with many of them being used for metrological purposes either to calibrate instruments, or forming the basis of measuring devices such as encoders. Evaluating the period of one or two-dimensional periodic structures from topography measurements, \emph{e.g.} performed using
scanning probe microscopy (SPM) methods, can be achieved using different 
methodologies with many grating evaluation methods having been proposed in the past and applied to a handful of examples. The optimum methodology for determining the grating pitch is not immediately obvious. This paper reports the results of extensive large-scale simulations and analysis to evaluate the performance of both direct and Fourier space data processing methods. Many thousands of simulations have been performed on a variety of different gratings under different measurement conditions and including the simulation of defects encountered in real life situations. The paper concludes with a summary of the merits and disadvantages of the methods together with practical recommendations for the measurements of periodic structures and for developing algorithms for processing them.
\end{abstract}

\vspace{2pc}

\section{Introduction}

Surface topography measurements are one of the important tools in the area of nanoscience and nanotechnology.
The more complex the properties that are measured, the more important is the selection of the data processing methodology, as surface topography measurements do not provide the result directly. In this paper we deal with evaluation of grating parameters and similar periodic structures, which can be used for a variety of applications.  
Periodic structures in the nanometre to micrometre range are frequently found in the areas of surface science, nanoscience and nanotechnology. Starting from the nanoscale, the atomic lattices, that are providing insight into the atomic arrangement of matter are periodic and their properties can be evaluated from very high spatial resolution measurements of surface topography\cite{Kim22,Li21}. The lattice parameter of silicon is even recognised as a secondary realisation of the metre for dimensional nanometrology \cite{Yacoot_2020_MEP} At larger scale, surface topography is being used to analyse artificially created 2D periodic structures that can be used as metamaterials\cite{Kondratov2017}, photonic crystals\cite{Romano18,Xiong19,Panfilova2019} or phononic structures\cite{MayerAlegre11}, controlling the way the energy passes through the matter. In all these cases the period is one of the key characteristics, directly affecting the material function.

When it comes to manufacturing methods themselves and related measurement science, there are also different potential roles for periodic structures. First of all, they can be used as key components in the manufacturing process, such as gratings used for mask 
overlay adjustment in the semiconductor industry\cite{Schmidt12}.
Second, they can be a critical part of the measurement device itself. Most of the area sensors, \emph{e.g.} CCD chips
are also periodic structures with an arrangement similar to 1D or 2D gratings and position of individual features may have direct
impact on the measurement accuracy, \emph{e.g.} when sub-pixel accuracy needs to be achieved, as in astronomy, and when both the geometrical errors and electronic performance defects of individual pixels need to be considered\cite{Peterson20}, or, at larger scale, when making accurate X-ray tomography measurements\cite{Luthi20}.  
Shack-Hartmann wavefront sensors combine a 2D periodic microlens array with a CCD chip, mounted in the focal plane of the microlens array (usually this is calibrated by using an ideal wavefront). Grating structures  are used as the basis of optical encoders and 2D gratings are being adapted for multi-axis position sensing\cite{Li13c}. Self-assembled periodic particle or hole arrays can be used as substrates for surface enhanced Raman scattering measurements\cite{MosierBoss17}, providing the plasmonic field enhancement. All the above mentioned structures need to be either measured or calibrated at some stage and one of the approaches is to use some of the surface topography measurement methods,
like scanning probe microscopy (SPM).
A major role of periodic structures is their use as transfer standards, providing metrological traceability for microscopes, thereby playing an important role in the traceability chain for dimensional metrology.
An important example of this application is related to the family of scanning probe microscopes, in particular atomic force microscopes (AFMs),  that have provided a gateway into the nanoscale world. AFMs have applications that include primarily imaging but also measurement of dimensions, surface roughness, electrical and magnetic  properties as well as chemical analysis and manipulation of structures and lithography. 
To make these techniques quantitative rather than qualitative and to be able to relate any measurements to the real world rather than just the microscope's frame of reference, accurate calibration of the scanning probe microscope is necessary. The ISO standard  \cite{ISO-11952}, \emph{ISO 11952 Surface chemical analysis --- Scanning-probe microscopy --- Determination of geometric quantities using SPM: Calibration of measuring systems},  describes the calibration procedure for scanning probe microscopes, where calibrated step height, pitch and flatness standards are required for complete instrument calibration.
Here, the grating serves as the pitch standard and therefore must be itself calibrated. This is normally done either by diffraction or using a metrological atomic force microscope \cite{Yacoot2011}. Optical diffraction has the advantages of being quick and having direct traceability to the metre via the wavelength of the light diffracted from the grating. The disadvantages of the method are the limitation of the diffraction limit preventing gratings with sub-wavelength pitch being calibrated and also the spot size of the light being used means that the method provides a global value for the pitch of the grating averaging out any local variations. On the other hand, measurements made using a metrological atomic force microscope overcome these issues, but data sets acquired require further processing to obtain the grating pitch. Evaluation of periodic structures from topography data is therefore needed for SPM calibration.





In order to evaluate periodic structures from topography measurements, one can use the ISO standard (11952)\cite{ISO-11952} designed for calibration of SPMs. This begs the question, why there is a need to revisit the calibration of pitch standards?  Experience shows that in the main, people predominantly use only the simplest possible methods, even if the ISO standard recommends more sophisticated methods. Based both on our experiences with SPM data processing software development over last 20 years
and on the experiences from an interlaboratory comparison\cite{Yacoot2020}, we see that
many SPM users evaluate grating period only from a distance of two crossing points on a line profile, or, base their uncertainty in the grating pitch solely on the  standard deviation, from multiple distance measurements. So, many users still struggle to calibrate their AFMs using gratings, thereby limiting their ability to make basic dimensional measurements.
Moreover, many users tend to systematically underestimate the area that is needed for obtaining
statistically significant results, as shown already also for roughness measurements\cite{Necas20_SR}.
The intuitive choice of the scan area is not optimal and without a detailed analysis it is hard to understand why this is the case. 

One of the reasons why people are struggling with periodic structures analysis might be 
the wide variety of choices available. There are many methods in the literature 
and standardisation documents that can be used for evaluation of periodic structures and their explanation is not always clear.
The Fast Fourier Transform (FFT) method, gravity centre method and a combined method (FFT+cross-correlation) are at present recommended by ISO 11952. Even though there is guidance on the number of grating periods that should be measured 
on the sample and suggested resolution (\emph{e.g.} more than 5 for gravity centre method and more than 7 for FFT methods),
there is no detailed analysis in the standard or referenced papers of the dependence of resulting uncertainty on the size of the area scanned.  It can be seen that preferably a large number of periods should be chosen for achieving the smallest uncertainties \cite{Dai05, Dai07}. However, it is hard to guess how the uncertainty 
will grow when these ideal conditions are not met, as is likely to be the case in reality.
When a grating is measured, the main limitations are the scanning range
and grating pitch. The scanning ranges can be varied as well as the choice of grating. The typical maximum scanning range for an AFM is to 100\,\textmu m. The range of most grating pitches is between 100 nm and 5\,\textmu m, for SPM calibration gratings, but can be very different in other cases. The typical number of pixels in an SPM image is is 500 to 5000 pixels. This gives users a wide choice of parameter combinations and from the AFM comparison \cite{Yacoot2020},  it was seen that users chose a variety of pitch/scan range parameters.
In this paper we consider the effect of this poor choice.
Moreover, it is not clear how the different aspects of non-ideal measurement
(feedback loop effects, noise, tip convolution) or non-ideal grating parameters (form errors, roughness) and of course limited scan range, affect the result. Special cases are the lateral imperfections of the grating such as misplaced grating pits or stitching errors introduced in the fabrication process. Even if the calibration gratings available on the market are very good from this point of view, the uncertainties
of present metrological SPMs are at the nanometre level so the measurements should also address this aspect.
Moreover, grating analysis methods are used for analysis  of other periodic structures used in industry
(CCD chips, microlens arrays) where the positioning error can be much larger.

There are different types of periodic structures that  could be evaluated using surface measurements. {Figure \ref{fig:periodic} shows synthetic images of some typical periodic structures, including some typical experimental errors and sample imperfections.

\begin{figure}
\includegraphics[width=\hsize]{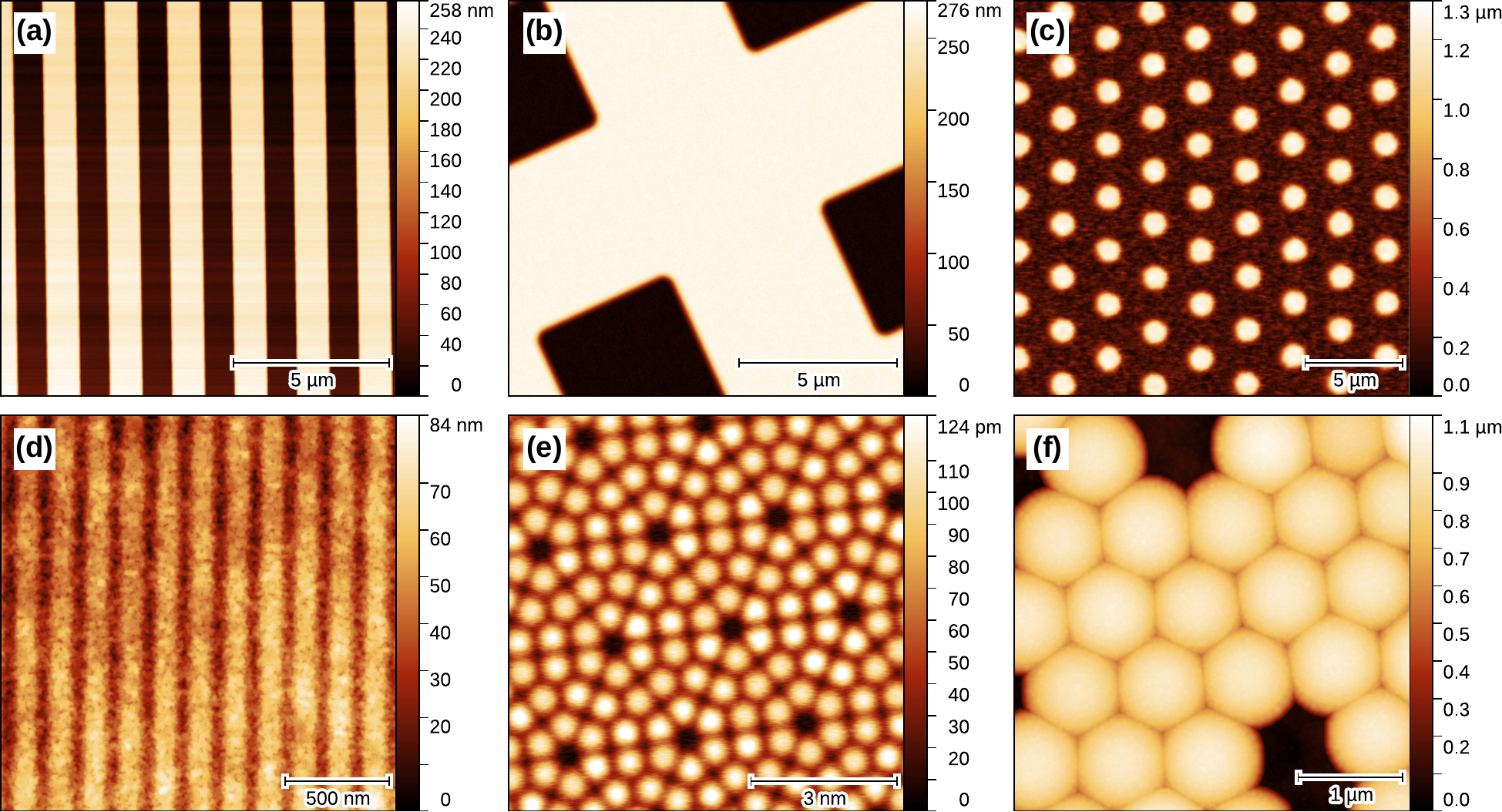}
\caption{Examples of periodic structures: (a) an ideal 1D grating; (b) a measurement of a too small an area on a 2D grating; (c) nanowires; (d) a poor quality 1D grating; (e) an atomic lattice --- Si $7\times7$ surface reconstruction; (f) self-assembled spherical particles.}
\label{fig:periodic}
\end{figure}

\section{Methods}

The goal of this paper is to systematically assess the performance of various implementations of the periodic structure evaluation methods. Therefore, the methodology is purely based on numerical simulations, using synthetic data with known
parameters \cite{Necas21}. Using the data synthesis tools in the open access AFM data analysis software Gwyddion \cite{Gwyddion}, grating surfaces with different properties and deterministic distortions can be generated. 
In detail, for all the simulations many gratings with slightly different parameters were generated in each simulation to suppress coincidences and aliasing effects. The inter-instance parameter variation range was 5\,\%. If, for example, the nominal grating period was 50 pixels (and this value would be shown in a figure), individual generated gratings would in fact have periods from 47.5 to 52.5 pixels.
Various data processing methods can be then applied to the generated gratings and their performance can be evaluated statistically.
In contrast to using real data this can directly provide a quantification of the errors as the true parameters of generated surfaces are known. It also permits a wider variation  in the value of parameters that would be possible with experimental work.
The analysis is performed initially on 1D gratings, to introduce the methodology and different error source types. 
Then, the differences obtained when evaluating 2D gratings are presented and discussed.\\

Methods suitable for analysis of gratings and other periodic structures that can be found in the literature 
and in ISO 11952 can be divided into the categories as presented below together with details of their implementation.
Eight period evaluation methods were implemented. Their principles are illustrated in \fref{fig:methods}.
\begin{figure}
\includegraphics[width=\hsize]{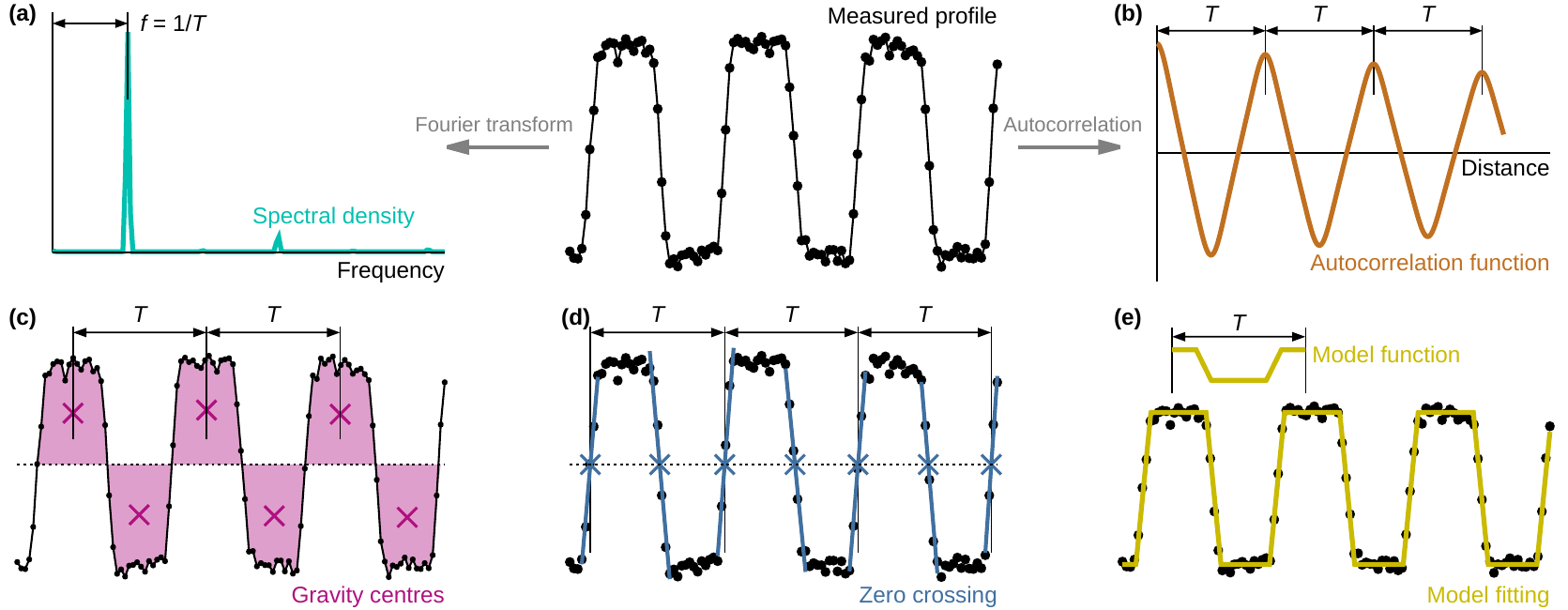}
\caption{Principles of methods for one-dimensional grating period evaluation: (a) Fourier transform based methods; (b) autocorrelation function based methods; (c) gravity centres; (d) zero crossing; (e) fitting of a model function.}
\label{fig:methods}
\end{figure}

\subsection{Direct space measurement}

As mentioned earlier, direct space measurement can be the most intuitive approach,
namely when a profile is drawn across the periodic structure and the lateral distance
between two or more similar features on the profile is evaluated\cite{Misumi2003, Huang2006a}.
More advanced approaches use more elaborate methods to determine the period, \emph{e.g.} by evaluating
the position of the profile at some height, interpolating or fitting the points near to this position to obtain the position more accurately. This is called the zero crossing method and is also often used\cite{Ortlepp2021,Chen2017,Huang2006a}. To use only points related to a special feature and throw away the rest of the structure might be not very efficient,
so the centre of gravity evaluated from whole structures can be used in order to include all of the data
\cite{Misumi2003a,Ortlepp2021,Dai07,Chen2017}

Two methods working in the direct domain were implemented for this study:
\begin{description}
\item[Gravity centre] (GC) --- analysis of gravity centres of grating bars \cite{Dai05,Chen2017}. The measured profile is plotted as a function, a threshold is chosen and the profile shifted to make the threshold line $z=0$. The centre of gravity of each bar is then defined as the center of gravity of the area under the curve (\fref{fig:methods}c). Its horizontal coordinate is $x_\mathrm{c}=\int xz\,\rmd x/\int z\,\rmd x$ (both integrals are over the bar interval). We obtain a set of centres $x_{\mathrm{c},n}$, indexed by integer bar number $n$, and fit them with a linear function
\begin{equation}
x_{\mathrm{c},n} = nT+c
\label{gc-fit}
\end{equation}
with parameters $T$ (period) and $c$ (offset). This can be repeated with areas above the curve for the negative parts and the results averaged.
\item[Zero crossing] (ZC) --- analysis of positions where the profile crosses the zero line \cite{Huang2006,Huang2006a,Chen2017}. A zero line is chosen as in the gravity centre method. Data around each zero line crossing are fitted with straight line $z=a+bx$ to estimate precisely the crossing coordinate $x_0$ (\fref{fig:methods}d). We obtain the coordinates of up-crossing positions $x_{0,n}$, indexed by integers $n$, and fit them with a linear function
\begin{equation}
x_{0,n} = nT+c
\label{zc-fit}
\end{equation}
with parameters $T$ (period) and $c$ (offset). This can be repeated with down-crossings and the results averaged.
\end{description}
The implementation of some methods is straightforward, whereas others require more care to work reliably. Some comments on our experience with their implementation are made after the results of numerical simulations have been presented.

\subsection{Fourier transform}

Next class of methods is based on the spectral density of spatial frequencies 
that can be obtained using Fourier transform (FT). A 1D or 2D discrete Fourier transform (DFT) is run on the measured
topography and from the frequencies corresponding to the peaks the period is evaluated\cite{Dixson2010,Ortlepp2021}. 
As the lateral size of the scan is limited, the frequency resolution can be very coarse,
which can be handled using different approaches for calculating a refined FT \cite{Dai07,Jorgensen1998},
\emph{e.g.} calculating the spectral density also for non-integer components.

The following methods working in the frequency domain were implemented for this study:
\begin{description}
\item[Naïve FFT] --- an elementary Fourier transform based estimation. The DFT of the measured profile data $z_n$ (both $n$ and $\nu$ take values 0, 1, 2, \dots, $N-1$)
\begin{equation}
Z_\nu = \sum_{n=0}^{N-1} z_n \exp\left(-2\pi\rmi\frac{n\nu}{N}\right)
\label{dft}
\end{equation}
is computed using an FFT. We then find the index $\nu$ where the spectral density $|Z_\nu|^2$ attains its maximum and take the corresponding spatial frequency $f=\nu/(Nh)$ as the pitch value, where $h$ is the sampling step. The period is obtained using the relation $T=1/f$ (\fref{fig:methods}a). Prior to the Fourier transform, data are multiplied by a windowing function. The simple raised cosine Hann window was used (in all frequency domain methods).
\item[Dai05 FT] --- refined Fourier transform \cite{Jorgensen1998,Dai05}. A coarse estimate is computed using the previous method and then refined by allowing non-integer values of $\nu$. The search for the precise maximum starts with the interval $[\nu-1,\nu+1]$ around the integer coarse maximum $\nu$. This interval is then progressively refined using a simple grid search, until it becomes shorter than a prescribed length. Fourier coefficients are computed by a direct evaluation of expression~\eref{dft}.
\item[Zoom FFT] --- Zoom-FFT refinement is, in principle, equivalent to the preceding method but computed in a different way. Fourier coefficients for non-integer $\nu$ are not computed individually; Instead Bluestein's algorithm~\cite{Bluestein1970} is used (discussed in more detail in \sref{sec:fft-refine}). It computes Fourier coefficients corresponding to an arithmetic progression of frequencies $f,f+\Delta f,f+2\Delta f,\dots,f+n\Delta f$ using an FFT. Therefore, it is possible to zoom into the interval around the coarse maximum and refine its position using a simple search. The results presented here show the effect of zooming in  the interval $[\nu-1,\nu+1]$, computing again $N$ Fourier coefficients, which corresponded to an  $N/2$-times refinement.
\end{description}

\subsection{Autocorrelation}

A direct domain parallel to the spectral density is the auto correlation function (ACF). The methods
implemented for this study can be considered hybrid as the measurement is made in the direct domain, but the ACF is obtained using the FFT:
\begin{description}
\item[Naïve ACF] --- an elementary autocorrelation-based estimation. The discrete ACF
\begin{equation}
G_k = \frac{1}{N-k} \sum_{n=0}^{N-1-k} z_n z_{n+k}
\label{acf}
\end{equation}
is computed using FFT, utilising the discrete cross-correlation theorem. Integer indices, $k$, are related to real distances $\tau$ via the sampling step $h$: $\tau=hk$. Its first maximum always lies at zero. The next one corresponds to the grating period $T$ (\fref{fig:methods}b) and is directly used as the estimate. 
\item[Multi-peak ACF] --- multiple ACF maxima are used to improve accuracy. The position of a single ACF maximum cannot be determined very precisely. However, the function has many maxima corresponding to integer multiples of $T$ (\fref{fig:methods}b). We locate as many of them as feasible, obtaining a set of horizontal distance $\tau_n$, indexed by integers $n$, and fit them with a linear function
\begin{equation}
\tau_n = nT
\label{multiacf}
\end{equation}
with a single unknown parameter $T$.
\end{description}

\subsection{Fitting}

Most of the methods used for grating analysis provide only pitch and angle as a result. There are, however, more parameters that could be evaluated on a grating; roundness of corners or fill ratio, that cannot be determined using the standard methods but still could provide some information for practical use of the grating, 
\emph{e.g.} when one wants to characterise the shape of the AFM tip using the grating. Although tip shape is not strictly necessary when evaluating grating periods, it would be useful if the sample had other features to be measured that were non-periodic. The grating height, if constant throughout the grating structure could be used for $z$ calibration 
and extracted background could, in principle, be used instead of basing it on separate measurements of a flatness standard. However, care should be taken when calibrating the $z$ axis as the optimum measurement strategies for pitch and height are different. 
Fitting the data using a model is used in many areas of measurement science and using this approach
for grating analysis could be understood as a straightforward approach. Fitting grating parameters
can be already done in SPM data processing software, such as Gwyddion \cite{Necas12} which has the capability
to evaluate many other parameters rather than just grating pitch.

For least-squares fitting, a suitable function describing the grating shape must be first chosen, for instance, a rectangular or sine wave. The function has several unknown free parameters: period $T$, height, offsets in $x$ and $z$, and possibly others such as slope width, describing the shape in more detail. An initial estimate of their values is necessary, for example this can be obtained using the simple FFT method. Precise values are then obtained by non-linear least-squares fitting of the model function using the Marquardt--Levenberg algorithm~\cite{More78}. In this study a rectangular wave with sloped walls was used as the model function (\fref{fig:methods}e).


\subsection{One-dimensional gratings}
\label{sec:methods-1d}

One-dimensional gratings were modelled as rectangular waves with slightly sloped walls (5\,\% of length). 
The following random and scanning artefact types could be added to the ideal grating data, individually or in combinations (a graphical illustration of the artefacts can be seen in \sref{sec:grating-1d} in \fref{fig:scaling}):
\begin{description}
\item[Waviness] --- the deviation of the grating substrate from ideally flat surface was added as a multi-scale locally smooth random additive background.
\item[Unevenness] --- the grating geometry imperfections were added using uncorrelated random variation of individual bar parameters, including position, height and fill ratio.
\item[Broken bars] --- another grating imperfection was introduced by random removal of top parts of individual bars (up to complete removal).
\item[Particles] --- the presence of dust particles was added using random bumps with size typically comparable to or somewhat smaller than one grating bar.
\item[Noise] --- the impact of SPM noise was added using independent random Gaussian noise of each sample.
\item[Tip convolution] --- the impact of SPM probe-sample convolution was added by convolution with an ideal parabolic tip.
\item[PID loop] --- the impact of the feedback loop imperfections was added using a simple proportional-integral-derivative feedback loop simulation.
\end{description}
Particles and broken bars were always used together as one `local defects' artefact. Other important systematic error sources exist: erroneous calibrations, drift, Abbe error and cosine error. They change the measured data to give a slightly different value for the grating period. Given such data, even a hypothetical ideal evaluation would compute the changed period. From the data processing standpoint they are, therefore, not interesting as all methods are affected in exactly the same way.

The evaluation also included a preprocessing step with two main goals: suppression of long-wavelength background (waviness) and shifting the profile mid-height to $z=0$. The latter is required mainly by the zero crossing method (ZC) and to a lesser degree the gravity centre method (GC), but the same preprocessing was used for all methods. Waviness was removed using a custom envelope method (see \sref{sec:background} for an explanation as to why it was chosen). Upper and lower profile envelopes were found as the local maximum and minimum within $1.5T$ interval (with $T$ estimated using Naïve FFT). Their average was processed using a low-pass Gaussian filter ($0.5T$) and subtracted from the data. The mid-height was located by finding the two main peaks in the height distribution and taking their midpoint~\cite{Ortlepp2021}.

\subsection{Two-dimensional gratings}
\label{sec:methods-2d}

Two-dimensional gratings were generated with slightly rounded rectangular holes, half the period wide, corresponding to 3/4 of upper surface and 1/4 of lower surface. The two lattice vectors could differ in length, but they were always orthogonal. The orientation and offsets (phases) in the plane were random. A typical image is shown in \fref{fig:grating}a. As in the 1D case, individual grating instances varied within 5\,\% of the nominal values.

The simulated images could include three artefacts analogous to 1D: noise, waviness and uneven positions. In addition, random tilt could be added as it is ubiquitous in AFM images. Images usually cover much smaller areas than scan designed to capture what is effectively a line profile, specified by the number of periods and several scans very close together perpendicular to the profile. Therefore, only this simple background was considered. Also the preprocessing was  simpler and only included tilt removal. It was implemented as an initial plane levelling, followed by splitting the surface to upper and lower portion using Otsu's threshold~\cite{Otsu79} and final plane levelling using only the upper portion of the surface (as defined by the threshold).


\begin{figure}
\includegraphics[width=\hsize]{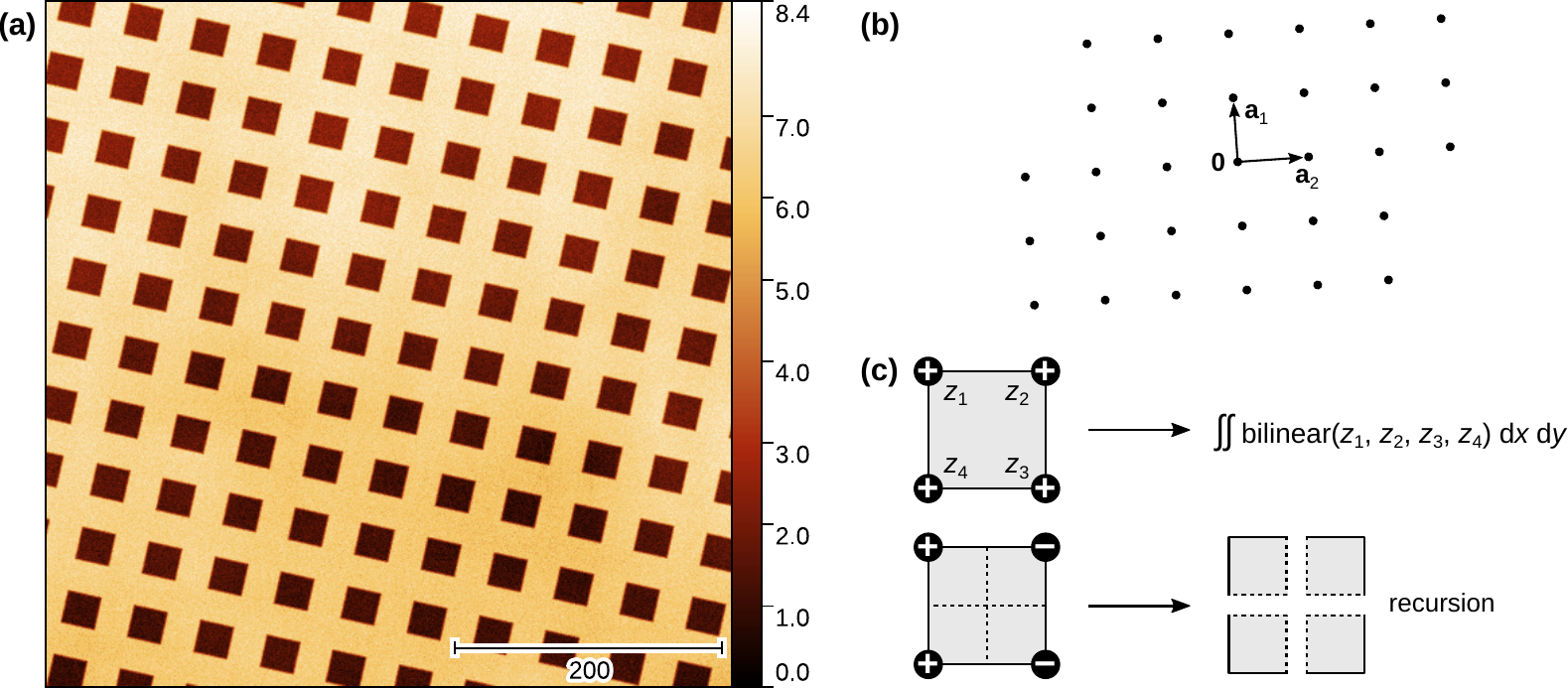}
\caption{(a) An example of typical grating image used in the simulation, with dimensions $500\times500$ pixels, period $T=50$ pixels and a combination of selected artefacts. All values are rescaled to pixels, \emph{i.e.} unitless. (b) Point lattice which is the usual intermediate step of 2D grating evaluation. (c) Recursion in the 2D GC computation.}
\label{fig:grating}
\end{figure}

A subset of 1D evaluation methods was implemented: naïve FFT and ACF, refined FT, multi-peak ACF, GC and model fitting. Zero crossing was not implemented because it is not clear how it generalises to 2D. Both FT and ACF based methods generalise directly to 2D. Refined FT was only implemented using Zoom-FFT as the two refinement methods are equivalent (this was already verified for the  1D case).

Both FT and ACF methods produce sets of points in a more or less regular lattice (\fref{fig:grating}b) and two lattice vectors have to be selected~\cite{Jorgensen1998}. The procedure can be outlined in two steps:
\begin{enumerate}
\item Find the point closest to the origin, but not at the origin. Use its position as lattice vector $\mathbf{a}_1$.
\item Find the point which is closest to the origin and linearly independent (sufficiently small scalar product with $\mathbf{a}_1$). Use it as $\mathbf{a}_2$.
\end{enumerate}
Integer indices of any other point $\mathbf{v}$ are then determined (where necessary) by solving $\mathbf{v}=m\mathbf{a}_1+n\mathbf{a}_2$ for $m$ and $n$ and rounding them to integers. Then the positions are fitted by linear least squares model similar to expressions \eref{gc-fit}, \eref{zc-fit} and \eref{multiacf}, only 2D:
\begin{equation}
x_{m,n} = ma_{1x}+na_{2x}\;; \quad y_{m,n} = ma_{1y}+na_{2y}
\end{equation}
with free parameters $a_{1x}$, $a_{1y}$, $a_{2x}$ and $a_{2y}$.

GC has two steps, identification of holes and computation of their centres. Otsu's threshold was again used to choose the $z=0$ plane and mark the holes. Holes touching image borders were filtered out as well as holes that were too small (single-pixel holes). The GC was defined exactly the same as in the 1D cases, even though its computation was more involved, requiring integration over the region where interpolated data lie below the $z=0$ plane. A simple recursive quadrature was used (\fref{fig:grating}c). We started with with pixel-sized rectangles formed by $2\times2$ neighbour values and then:
\begin{enumerate}
\item If the values at all rectangle corners were negative it was considered completely covered. The integral over the rectangle was computed analytically using the bilinear interpolation of the four corners.
\item If all corners were negative the rectangle was skipped.
\item If some corners were positive and some negative, the rectangle was split into four, with corners computed by interpolation, which were then evaluated recursively  .
\item If a rectangle became too small the recursion was terminated.
\end{enumerate}
The image has $N^2$ pixels but only $O(N)$ are at hole boundaries, requiring recursion. Therefore, a more efficient integration method was not necessary. After finding all centres, one close to image centre was chosen as the initial origin. The analysis then proceeded as for the case of a multi-peak ACF, except that the origin position was also a free parameter and it was updated after each fitting step.

Model fitting was implemented as in the 1D case, again relying on naïve FFT for the initial lattice vector estimates. The model function was similar to the grating generation function. However, it was more general, allowing non-orthogonal lattice vectors. Not doing so would give fitting an unfair advantage over the other methods since the generated gratings had orthogonal lattice vectors.

\subsection{Gratings with small number of periods}
\label{sec:method-small}

Data with just a couple of periods are evaluated differently than data with a thousand. The profile is likely to be scanned more slowly with respect to feature size, the entire data can be inspected and ensured they are defect-free, levelling and zero line can be checked manually. It is not entirely fair to use the models and methods outlined in sections \ref{sec:methods-1d} and~\ref{sec:methods-2d}, focused on automated processing of long profiles and large areas, to study this case. Therefore, they were modified for the 1D case as follows:
\begin{itemize}
\item The grating model was a perfect rectangular wave, distorted by the convolution with a rounded triangular tip (which was the only systematic artefact considered).
\item There would be no significant local defects and the background would be mainly  tilt, corrected by the user. Hence, we considered only one random artefact: noise.
\item The zero line was set exactly at mid-height.
\item A `manual' evaluation method was included.
\end{itemize}
The typical choice of two points for manual measurement is zero crossings, so that is what the `manual' method used (it was still carried out automatically). The two most distant zero crossings of the same type (up or down) were found, located with subpixel precision by linearly interpolating the two adjacent points, and their distance was divided by the number of periods between.

Other methods were tweaked to stretch their applicability. Usually, in GC and ZC both `up' and `down' features are analysed and the results averaged. This was still done when possible, but if the method could find at least one measurable $T$, the evaluation was considered successful. Zoom FFT started the initial coarse estimate from $3\times$ zoomed FFT, instead of plain FFT, because the peak could be indistinguishable from the peak at origin otherwise.

\section{Results and discussion}
\label{sec:results}

\subsection{Methods performance on one-dimensional gratings}
\label{sec:grating-1d}

\begin{figure}
\includegraphics[width=\hsize]{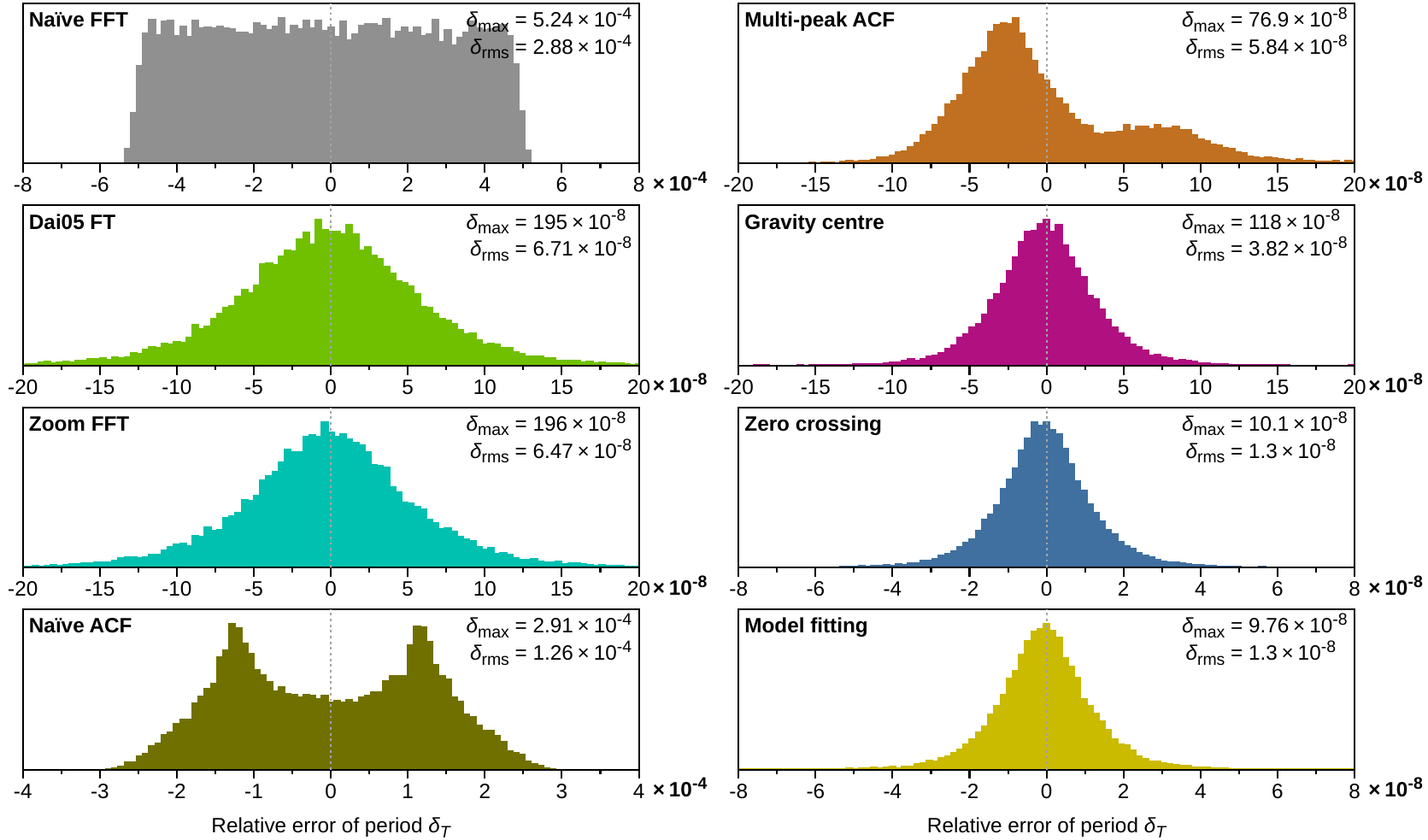}
\caption{Distributions of relative errors of the evaluated pitch for almost ideal gratings with approximately 1000 periods.  Note the order-of-magnitude factors in the bottom right corners of the graphs --- the errors can differ by several orders.  For some methods errors much larger than typical were occasionally encountered (as indicated by the maximum error $\delta_\mathrm{max}$).  In these cases the abscissae do not cover the outliers and corresponds roughly to the mean square error $\delta_\mathrm{rms}$.}
\label{fig:delta-distrib}
\end{figure}

The distributions of relative errors $\delta_T$ of the period are illustrated \fref{fig:delta-distrib} for a typical calibration grating with 1000 periods, 50 samples per period and waviness background with noise to signal ratio of 7\,\%  (approximately corresponding to \cite{Dai05}, by visual comparison). The distributions were obtained by running the evaluation on 25000 random grating instances. The error distributions are anything but Gaussian. Naïve FFT has a uniform error distribution, which is expected because the error is basically a rounding error. Naïve ACF has an odd bimodal error distribution, which is probably related to the use of parabolic interpolation to improve the maximum location. Its shape seems partially preserved in multi-peak ACF error distribution, which is also rather asymmetrical. All the seemingly Gaussian distributions have in fact heavier tails. The error distributions become more conventional when waviness is replaced by simple noise --- for instance multi-peak ACF asymmetry disappears. However, most simulated artefacts led to odd error distributions and heavy tails.

Both less accurate methods, naïve FFT and ACF, consistently give results with bounded errors; the maximum error encountered is a small multiple of the mean error. The more accurate methods occasionally give a value with much larger error than typical, causing the heavy tails of the distributions. For FFT-based methods this occurs when the period is very close, but not exactly equal, to an even number of pixels. Grating edges can then align with sampling points in a way that makes edge positions less certain than if there was no relation between the grating period and sampling step. Direct space methods are affected for a similar reason. In the following we will refer to the six more accurate methods as `the good methods' for brevity.

\begin{figure}
\includegraphics[width=\hsize]{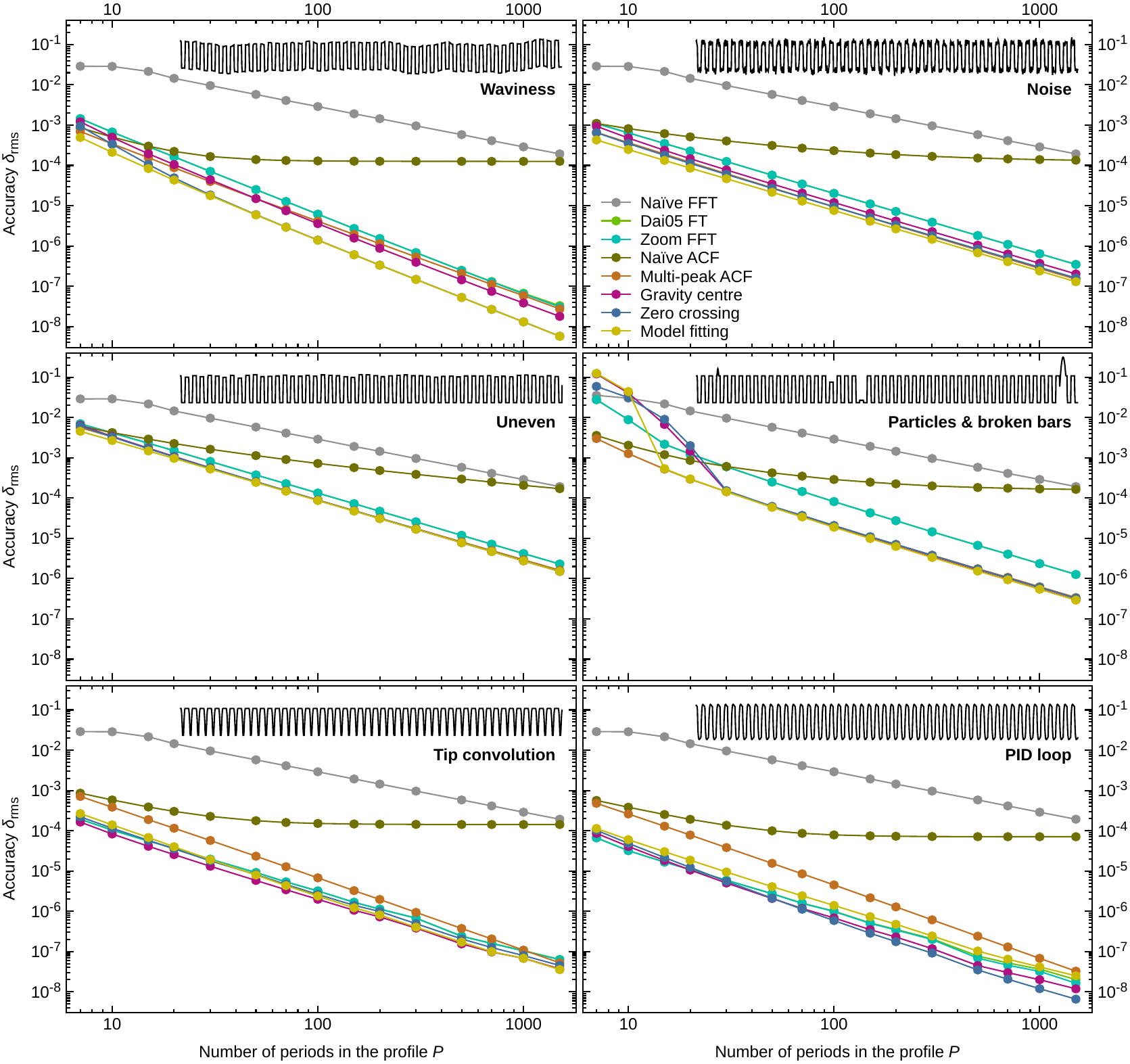}
\caption{Scaling of the accuracy (mean square relative error) $\delta_\mathrm{rms}$ of grating period with the number of measured periods $P$. Curves for the two refined FT methods cannot be visually distinguished. Each plots illustrates the corresponding type of artefact with features disturbed in scale to the simulation.}
\label{fig:scaling}
\end{figure}

\Fref{fig:scaling} compares how the methods behave when the number of periods $P$ in the profile change. For additive disturbances the noise ratio was kept at 7\,\%, likely to be the worst case encountered in a real situation. For uneven bars the relative standard deviation of parameters was 2.5\,\%. Particles and broken bars covered each randomly and independently 2\,\% of the profile. Tip convolution and PID loop are non-random effects and their parameters were chosen to obtain roughly comparable disturbance of the profile shape. The accuracy $\delta_\mathrm{rms}$ is measured as the mean square relative error.

There are some differences in sensitivity to different artefacts. Multi-peak ACF seem the most difficult to thwart by local defects (particles \& broken bars), but it is more susceptible to asymmetry of the shape (PID loop) than others. The rapid deterioration of accuracy of GC, ZC and fitting in the case of local defects and low $P$ is driven by occasional cases when an unfortunate constellation of defects manages to derail the method entirely. In most cases the accuracies of the good methods are comparable, although the logarithmic scale in \fref{fig:scaling} is deceptive as curves that appear close to each other can still differ by factor 2 or 3.

\subsection{Super-linear scaling}

\begin{table}
\caption{\label{tab:scaling}Estimated accuracy scaling exponents for 1D methods and different artefact types.}
\begin{indented}
\item[]\begin{tabular}{@{}lllllll}
\br
Method&Waviness&Noise&Uneven&Defects&Tip conv.&PID loop\\
\mr
Naïve FFT&1.00&1.00&1.00&1.00&1.00&1.00\\
Dai05 FT&1.97&1.50&1.50&1.56&1.49&1.46\\
Zoom FFT&1.99&1.50&1.50&1.56&1.50&1.51\\
Naïve ACF&0.05&0.27&0.57&0.32&0.09&0.13\\
Multi-peak ACF&1.86&1.54&1.49&1.54&1.79&1.81\\
Gravity centre&2.00&1.52&1.51&1.57&1.51&1.57\\
Zero crossing&2.05&1.51&1.51&1.56&1.55&1.73\\
Model fitting&2.05&1.50&1.50&1.57&1.61&1.54\\
\br
\end{tabular}
\end{indented}
\end{table}

For most methods, the accuracy in \fref{fig:scaling} clearly follows a power law. The scaling exponents are listed in \tref{tab:scaling}. As expected, the accuracy of naïve FFT scales linearly with $P$, the number of periods. Naïve ACF shows almost almost no improvement with an increased number of sampled periods beyond a certain point. It is limited in precision by the sampling inverval $h$ which does not change. The most interesting observation, however, is that the accuracy of all the good methods scales super-linearly with $P$. The exponent somewhat varies among them; it also varies somewhat with simulation settings. However, it is consistently around 3/2 or larger.

Several factors contribute to the super-linear scaling. The easiest case to analyse is the multi-peak ACF. The period is obtained by least-squares fitting of peak positions with the model equation \eref{multiacf}. Assuming for simplicity uniform uncorrelated Gaussian errors $\sigma_\tau$ of positions $\tau_n$, the estimated period is $\hat{T}=S_{n\tau}/S_{nn}$ and its variance $\mathrm{Var}[\hat T]=\sigma_\tau^2/S_{nn}$, where $S_{nn}=\sum_n n^2$ and $S_{n\tau}=\sum_n n\tau_n$. The sum $S_{nn}$ is just the sum of squares of the first $P$ natural numbers $S_{nn}=P(P+1)(2P+1)/6\approx P^3/3$ (for large $P$). Therefore, the standard deviation of $T$ is $\sigma_T\approx \sqrt3\,\sigma_\tau/P^{3/2}$, giving scaling exponent $3/2$.

The same effect is in play in GC, ZC and model fitting. If the ideal profile is disturbed by adding Gaussian noise, position errors are close to uncorrelated and normally distributed, all three indeed scale in line with the theoretical expression. In fact all the good methods scale similarly with exponent $\approx3/2$ in the white noise case. Deviations from 3/2 are, therefore, influenced by spectral properties of the disturbance as they scale non-trivially with the number of data points. For instance, the simulated waviness has a certain frequency spectrum and effectively disturbs short profiles more than long ones.

In refined FT methods the scaling comprises two factors. A factor $P$ comes with the increased frequency domain resolution, as in unrefined FFT. The additional $P^{1/2}$ is determined by how precisely the peak can be located inside one frequency step, \emph{i.e.}\ how deep we can `zoom'. Intuitively, the  DFT concentrates all direct space data to a few peaks in the spectrum. The peak width (measured in DFT frequency steps) does not depend on the profile length. So for longer profiles more data contribute to one peak and thus better define its shape.

\subsection{Optimal number of samples per period}

The super-linear scaling leads to an important conclusion concerning measurement strategies. Assume we measured a profile with $P$ periods. Now we want to increase the precision by measuring $5\times$ more data. Keeping the sampling step $h$, we can either repeat the measurement five times or measure a single $5\times$ longer profile. Although the first option is useful if representativeness can be an issue, the second is vastly more precise.  Measuring five times and averaging reduces the standard deviation by factor $\sqrt5\approx2.2$. However, measuring a five times longer profile reduces the standard deviation by factor $5^{3/2}\approx11.2$ for scaling exponent $3/2$.

Consider now that the maximum number of samples is limited, but the sampling step can be chosen freely. Should we measure many periods, a few, or is there a Goldilocks zone? If $\sigma_T\propto\sigma_x/P^{3/2}$ and the position error $\sigma_x$ is proportional to sampling step, \emph{i.e.}\ $\sigma_x\propto h\propto P$, it can be seen that $\sigma_T\propto1/\sqrt{P}$. Therefore, it seems as many periods as possible should be crammed into to the measured profile because the error decreases monotonically with $P$.

This suggestion may sound counter-intuitive. When performing SPM measurements, one often cares about pixel size, assuming that it is the key parameter limiting the accuracy of the result. However, in contrast to simple manual evaluation, all the presented methods are substantially sub-pixel by nature, by using all the available data.

\begin{figure}
\includegraphics[width=\hsize]{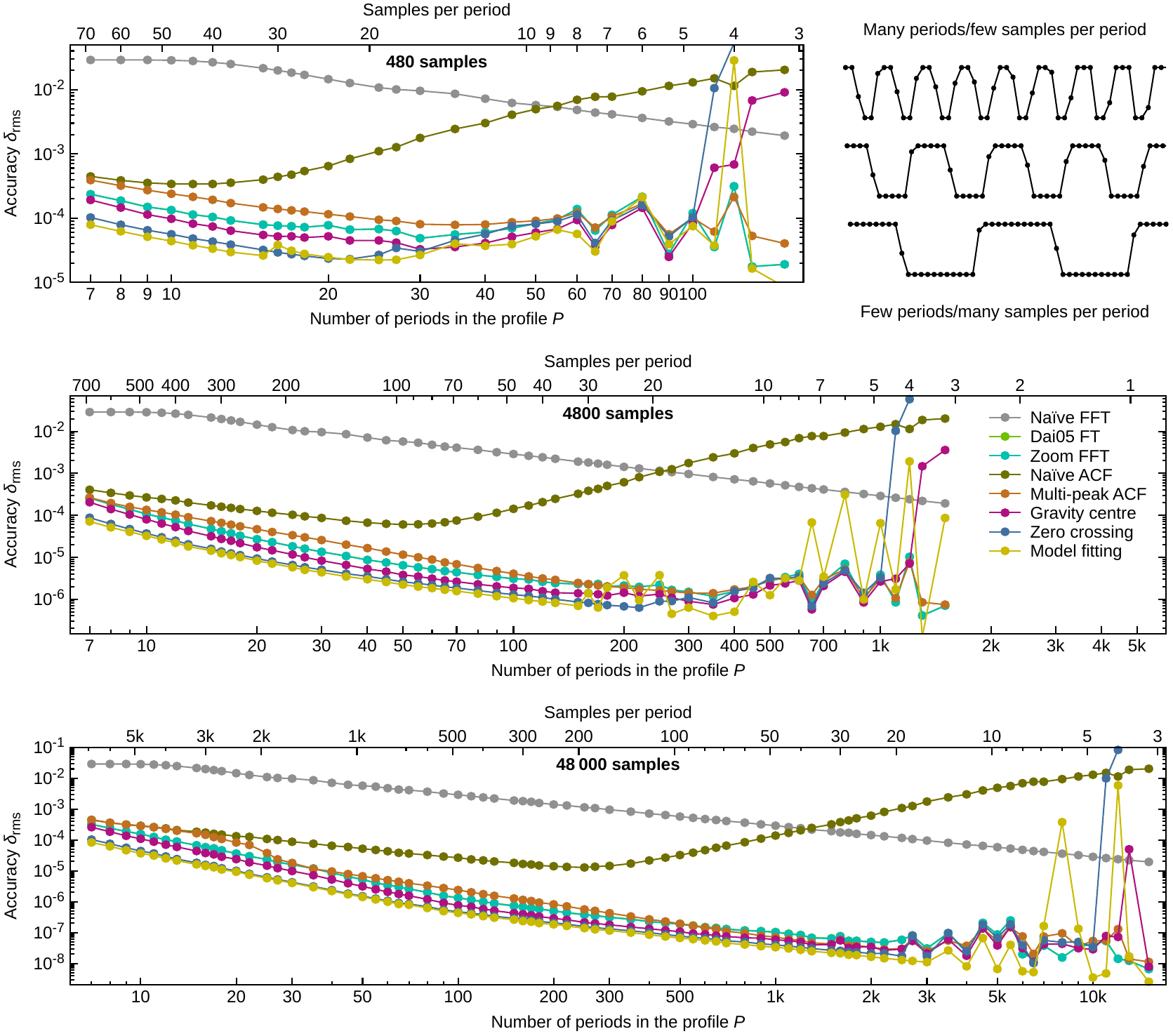}
\caption{Optimisation of the number of samples per period for 1D gratings. The mean square error $\delta_\mathrm{rms}$ of grating period is plotted as a function of the number of measured periods $P$ for three fixed numbers of samples in the profile (480, 4800 and 48\,000).}
\label{fig:optimum}
\end{figure}

Of course, if the sampling step becomes too long, and in particular when reaches or exceeds $T/2$, individual bars become impossible to distinguish and the scaling relation breaks. Therefore, there will be an optimal sampling step which was estimated  using a numerical simulation. Its results for fixed $N=480$, 4800 and 48000 and varying number of periods $P$ (that we again assume can be chosen freely) are shown in \fref{fig:optimum}. The three cases roughly correspond to a standard profile (possibly read from an image) using routine settings for pixel resolution on a commercial instrument, a profile obtained using commercial AFM at the limits of the possible pixel resolution, and a measurement using a specialized long-range metrological AFM capable of producing scans of virtually unlimited pixel resolution at the cost of low speed. A combination of random artefacts was used in this simulation, noise and waviness, each with 1\,\% noise, and slightly uneven bar parameters with $10^{-5}$ relative standard deviation.

All the good methods improve down to about 20--30 samples per period ($N/P$) before the errors level off and then become erratic, varying depending on how the sampling step and period align. The simple methods behave differently. For naïve ACF the optimum is $P\approx\sqrt{N}$. Naïve FFT improves steadily up to 3-4 samples per period. However, neither reaches the accuracy of the good methods. The optimum of 20--30 samples per period may depend on the defects present in the data. Nevertheless, it seems quite consistent over two orders of magnitude of $N$. It also agrees with the study of scanning speed influence~\cite{Ortlepp2021} where the profile length was kept fixed and $N$ decreased with increasing scanning speed. The variance of results did not change much for more than 20 samples per period, but it started to increase sharply when less than 20 points were measured (one needs to combine tables 1 and 2 in \cite{Ortlepp2021} to compute $N/P$).

To summarise, measuring more periods is better than measuring each position more precisely --- provided the sampling does not become too coarse and some other error does not grow too large. Still, the optimum settings can be quite counter-intuitive. This is a similar situation to roughness measurements where a measured area which `feels right' is often way too small~\cite{Necas20,Necas20_SR}. If we are limited by the maximum profile length, measuring with a shorter sampling step is still useful, but the precision gain is slower.

\subsection{Methods performance on two-dimensional gratings}

\begin{figure}
\includegraphics[width=0.5\hsize]{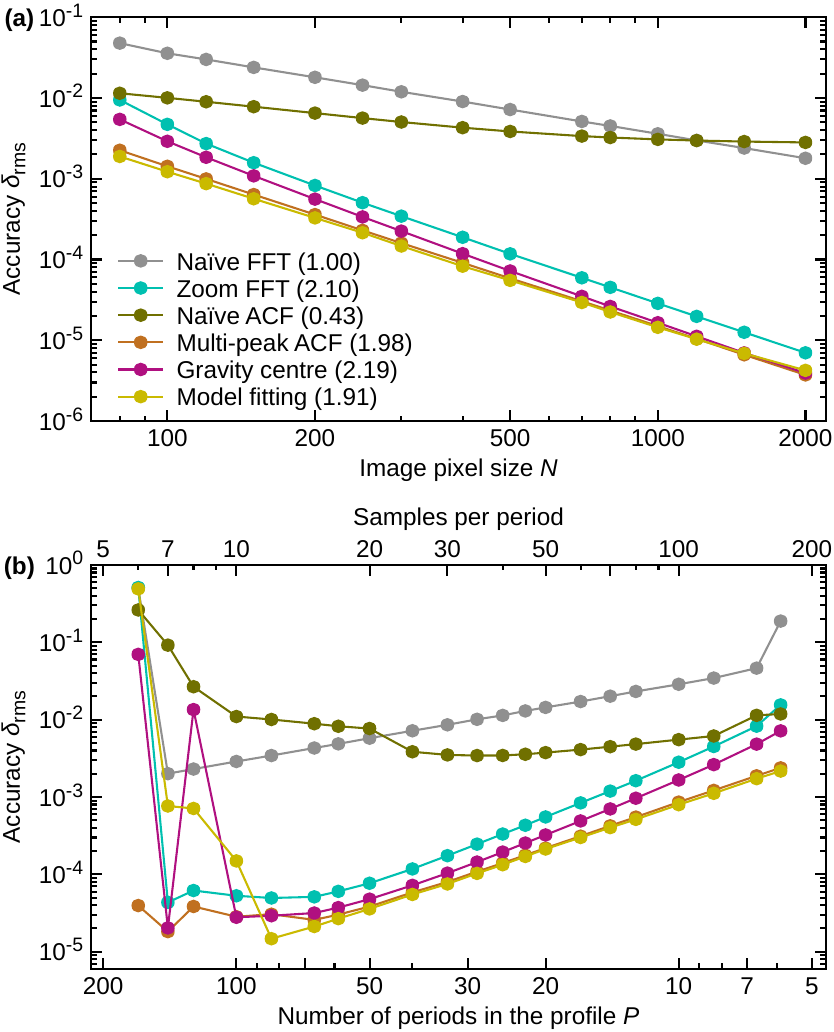}
\caption{Scaling and optimisation for 2D gratings: (a) accuracy scaling with image size, with estimated scaling exponents listed in the legend; (b) accuracy as a function of the number of measured periods for fixed image size $1000\times1000$ pixels.}
\label{fig:results-2d}
\end{figure}

The overall accuracy of the two lattice vectors $\mathbf{a}_1$ and $\mathbf{a}_2$ was measured as
\begin{equation}
\delta^2 = \frac{|\hat\mathbf{a}_1-\mathbf{a}_1|^2}{|\mathbf{a}_1|^2} + \frac{|\hat\mathbf{a}_2-\mathbf{a}_2|^2}{|\mathbf{a}_2|^2},
\label{delta2D}
\end{equation}
a natural extension of the 1D relative error to the 4D space formed by $\mathbf{a}_1$ and $\mathbf{a}_2$.

Results for 2D gratings are plotted in \fref{fig:results-2d} for combined random artefacts (noise + waviness + uneven + tilt). The scaling exponents for all four good methods are around two. They vary slightly, but this is expected based on the 1D results. The abscissa in \fref{fig:results-2d}a is the linear image size in pixels instead of number of periods $P$ used in \fref{fig:scaling}. An effective $P$ can be defined
\begin{equation}
P = \sqrt{\frac{|\mathbf{a}_1\times\mathbf{a}_2|}{h_xh_y}}\;,
\label{Peff2D}
\end{equation}
where $h_xh_y$ is the area of one pixel with sides $h_x$ and $h_y$. In the typical case $|\mathbf{a}_2|\approx|\mathbf{a}_1|$ and $h_y=h_x$. However, image size is easier to imagine.

Following the analysis of 1D scaling, the explanation of the scaling exponent is simple. Two vectors need to be determined. The accuracy of each vector scales with $P_\mathrm{along}^{3/2}$ for the number of periods along its direction $P_\mathrm{along}$. It also scales with $P_\mathrm{across}^{1/2}$ for the number of periods across its direction $P_\mathrm{along}$ because sufficiently distant profiles across behave like independent measurements. For a square image $P_\mathrm{along}\approx P_\mathrm{across}\approx P$. Therefore, the overall accuracy scales with
\begin{equation}
P_\mathrm{along}^{3/2} P_\mathrm{across\vphantom{l}}^{1/2} \approx P^{3/2}P^{1/2} = P^2\;,
\end{equation}
which is proportional to the number of image pixels $N^2$. Scaling with $P^2$ is again much better than with $\sqrt{P^2}=P$ which would follow from simply measuring more data. In higher dimensions $D$ the expected scaling is with $P^{D/2+1}$. In the 1D case we commonly saw mean relative errors below $10^{-6}$ with 50\,000 samples, whereas here we do not reach them even with $2000\times2000=2\,000\,000$ image pixels (for comparable noise levels). This is a direct consequence of the smaller number of periods in images and slower scaling with $P$. The error \eref{delta2D} is also larger because it is the total relative error of four vector components, instead of a single parameter $T$.

These observations together support the measurement strategy which converts 2D evaluations to 1D evaluations~\cite{Dai05}. The two lattice vectors are first found using a 2D measurement. They are then improved by measuring long thing stripes along each vector and evaluating them separately as this maximises the number of periods measured along each vector. Of course, this strategy can only be realised with a long-range metrological AFM.

As in one dimension, we can ask what is the optimal number of samples per period, measured using the effective $P$ defined by equation \eref{Peff2D}? The result, analogous to \fref{fig:optimum}, is plotted in \fref{fig:results-2d}b for images with $1000\times1000$ pixels. The overall behaviour is similar to the 1D case. The optimal number of samples per period for the good methods is lower than in 1D, about $N/P=10$. A possible reason is that the number of samples per grating unit is $(N/P)^2$, not $N/P$.

\begin{figure}
\includegraphics[width=0.5\hsize]{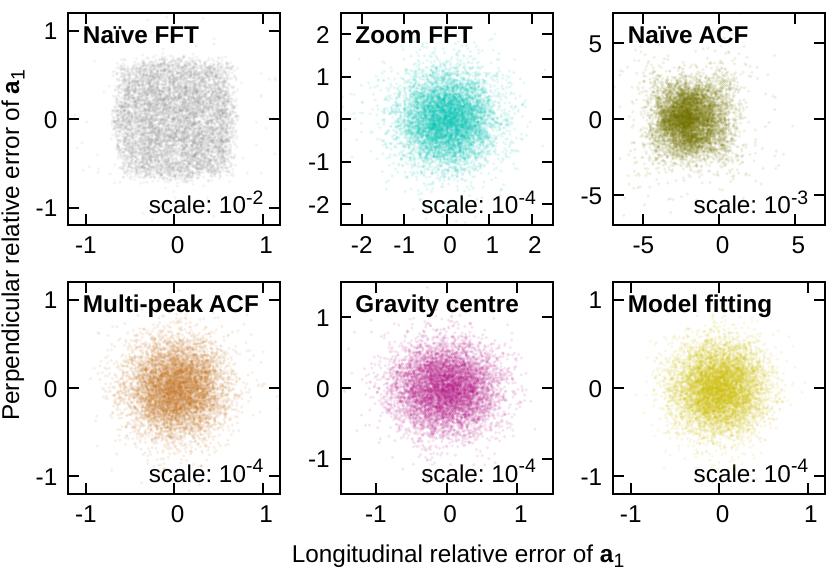}
\caption{Distributions of relative errors of 2D lattice vector $\mathbf{a}_1$ along the vector and in the perpendicular direction.}
\label{fig:distributions}
\end{figure}

Even though the generated hole patterns were always perfectly orthogonal, the evaluation did not impose any orthogonality constraint. The measured angle between $\mathbf{a}_1$ and $\mathbf{a}_2$ could thus deviate from $90^\circ$. In \fref{fig:distributions} we can see the 2D error distribution of $\mathbf{a}_1$ for $500\times500$ images. The naïve ACF method underestimates the length of $\mathbf{a}_1$ slightly and the error distribution for naïve FFT resembles more a uniform distribution in square than a Gaussian distribution. Distributions for the good methods are isotropic and the errors of angle and length can be thus estimated using the simple error propagation rule.

2D measurements are also used for 1D structures. In principle we then have two options, process the entire image using a 2D method, or extract a set of profiles and evaluate them using a 1D method. Only frequency domain methods and model fitting work identically as in the pure 1D and 2D cases. The rest are not directly applicable to images of 1D structures. As for profiles, they have to be taken along the lattice vector to prevent cosine errors, creating a chicken and egg problem since the lattice vector is what we are trying to determine. Even though the vector can be estimated using 2D FFT, if we are to compute 2D FFT a sensible strategy is to employ a refined FT method in 2D instead of returning back to profile extraction.

The analysis of scaling with image size remains unchanged from the pure 2D case. Now there is only one lattice vector, but the analysis considered each vector separately anyway. One may think that a $1024\times1024$ image would be equivalent to 1024 independent profiles, so the standard deviation would be reduced by factor $\sqrt{1024}=32$ compared to a single 1024-point profile. Although this can be the case, often such estimate is too optimistic. Individual scan lines and artefacts in them, such as line roughness, can be highly correlated and each line thus adds less independent information than the simple estimate suggests. This is analogous to roughness measurement which is plagued by the same problem~\cite{Zhao00,Necas20,Necas20_SR}.

\subsection{Small number of periods}

Two main reasons for measuring a small number of periods are instrumental limitations, \emph{i.e.}\ scanner range, and the preconception that it is necessary to measure the grating in fine detail for accurate results. We hope this work helps to dispel the latter, but the former is much harder to deal with. Measurements of short profiles/small areas are, and will be, common as most AFM scans are limited to \SI{100}\um{}.

A profile cannot be shorter than one $T$ ($P=1$) to measure the period. The manual method or ZC in principle require profiles only slightly longer than $T$, whereas GC needs profiles longer than $3T/2$ ($P=3/2$) to find two gravity centres of the same type. FFT methods are limited by the ability to distinguish the correct peak from the one at origin. Model fitting is interesting, in particular in 2D, by its ability to utilise information which is not along the two lattice vectors. The image in \fref{fig:periodic}b would not be the best measurement of the grating, but despite being `too small' and not accommodating measurements along the two lattice vectors, it can still be easily evaluated by model fitting (admittedly, the example is a bit contrived in order to illustrate the point).

1D simulation results are shown in \fref{fig:small}. The profile always had 1000 samples while $T$ and $P$ were varied. The additive noise was relatively low, 0.7\,\%. Since direct space methods may not find any usable points, the figure shows the accuracy $\delta_\mathrm{rms}$ for successful evaluations and also the success rate. Success was defined as (a) the method itself did not report failure, and (b) the result was not an obvious outlier.

As predicted, the success rate of GC drops rapidly below $P=2$, whereas other methods can work closer to $P=1$. Frequency domain methods never failed according to the criterion, but of course their accuracy is poor for a small $P$. The manual method appears to work well up to about $P=5$, at least in the low-noise case when two points suffice to find the intersection precisely. It is, in essence, a worse version of ZC, but for just a couple of periods they can behave similarly. Still, already in the range $P=5$ to 10 it is clear that the accuracy of the manual method remains constant, whereas the good methods start to scale with $P^{3/2}$ (as in \fref{fig:scaling}). We must stress that even in this case all the methods were most accurate for the \emph{coarsest} sampling (large $P$), not the finest sampling step.

Representativeness is a major concern when only a couple of features are measured. For periodic structures such as those in \fref{fig:periodic}c or \ref{fig:periodic}f the position and shape variation of individual features can be much larger than the precision with which we can measure the lattice vectors. Measurements at many different locations and statistical analysis are then essential.

\begin{figure}
\includegraphics[width=0.5\hsize]{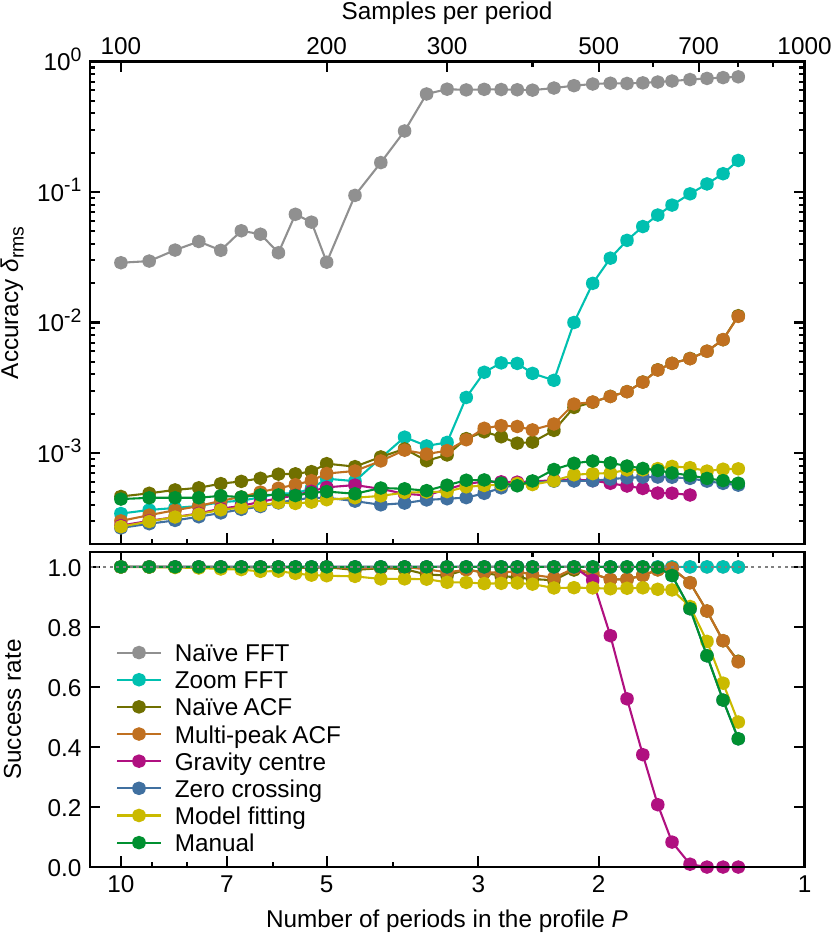}
\caption{Results for small number of periods.}
\label{fig:small}
\end{figure}

\subsection{Error estimates}
\label{sec:errors}

\Fref{fig:scaling} demonstrated that different artefacts with similar noise to signal ratios can result in order-of-magnitude differences in accuracy. It is, therefore, difficult to assign a universally valid accuracy estimate to each method. Frequency-domain methods do not directly give error estimates, but direct-domain and hybrid do.

GC, ZC and multi-peak ACF employ linear regression to obtain $\hat T$. In GC and ZC individual values can be considered independent and have finite variances. Therefore, the estimate of result standard deviation $\hat\sigma_T$ from linear regression should be usable. In ACF all the peaks are computed from the same data, so the independence assumption may not be justified.

Model fitting using non-linear least squares also gives standard deviation estimates $\hat\sigma_T$. They should be correct if data are disturbed by uncorrelated noise, although this is rarely the case. For large correlated artefacts the situation is more complicated and the standard deviations can be severely underestimated. They are also affected by systematic differences between the model and experimental data. 

The estimates provided by these four good methods were compared with the true error. The simulation was run for 1D profiles (4800 samples), $T$ of 50 pixels, and disrupted by the combined artefact (noise + waviness + placement). The noise to signal ratio was varied over a few orders of magnitude to obtain different $\delta_\mathrm{rms}$ and for each ratio 12\,000 grating instances were evaluated.

The results are plotted in \fref{fig:errors}. For all the methods the mean estimated relative standard deviation matched the true $\delta_\mathrm{rms}$ quite well. Only multi-peak ACF overestimated the error somewhat when it became small. For a large number of degrees of freedom the quantity $(\hat T-T)/\hat\sigma_T$ should be normally distributed. Inspection of the distributions showed that even though this was not entirely true, the distributions were not far from standard normal. The estimates from individual methods thus seem usable, with caution. Comparison of several different methods can also be helpful~\cite{Ortlepp2021}, despite not being directly usable for standard deviation estimates if we do not know how the errors are correlated.

\begin{figure}
\includegraphics[width=0.5\hsize]{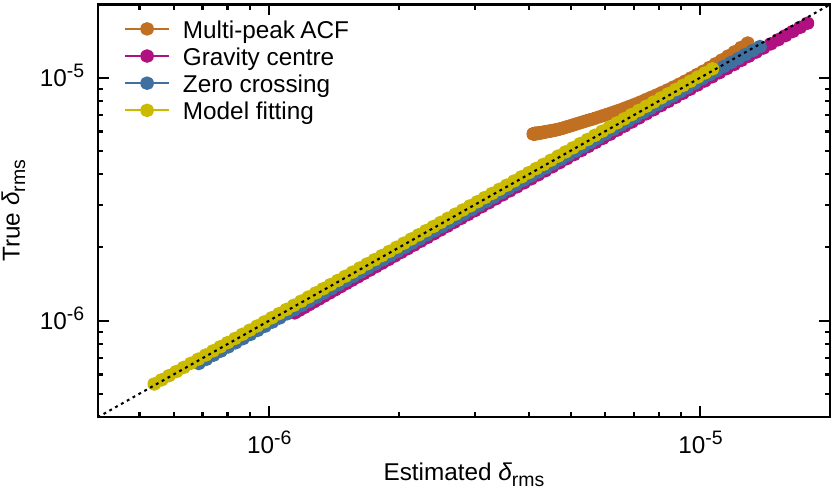}
\caption{Comparison of mean square relative error estimated from the method with the true error. The dashed line corresponds to the errors being equal.}
\label{fig:errors}
\end{figure}

Usually the accuracy is limited by artefacts in the experimental data and calibration uncertainties. However, one can also ask if the evaluation method can ever become the limiting factor  and at what accuracy level? The two naïve methods are obviously limited by frequency and sampling steps. For the good methods the answer is more interesting. It can be, unfortunately, also implementation-dependent. In order to investigate these intrinsic errors our implementations were run on ideal data. With no defects, two parameters remained to be chosen at random, the true period $T$ (within 5\,\% of nominal value, as usual) and grating phase.

The results are summarised in \tref{tab:ideal}. Naïve FFT and ACF behave as expected. For 1D data, ZC and model fitting could achieve more or less the full double precision, \emph{i.e.}\ they were limited by rounding errors. Interestingly, GC did not behave as expected, most likely because of subtle discretisation errors. They have been studied in detail under different assumptions than hold in SPM~\cite{Landi2002}, but are present also here. Other good methods also behave more or less similarly to GC. Peak apex locations in frequency domain methods can be subtly affected by windowing. For 2D data, model fitting did not achieve the same precision as in 1D and was not even improving much with image size. Although a portion of fit results was exact or within rounding errors, not all were --- unlike in 1D. The reasons are not clear; possibly the Gwyddion fitter has convergence problems for huge data sets. 

\begin{table}
\caption{\label{tab:ideal}Upper accuracy limit of individual methods, measured as $\delta_\mathrm{rms}$ for ideal data. Profiles had nominal period 50 samples; images 25 pixels.}
\begin{indented}
\item[]\begin{tabular}{@{}lccccc}
\br
Method&Profile 480&Profile 4800&Profile 48\,000&Image 250&Image 1000\\
\mr
Naïve FFT     &$3.1\times10^{-2\phantom{0}}$&$3.0\times10^{-3\phantom{0}}$&$3.0\times10^{-4\phantom{0}}$&$1.4\times10^{-2\phantom{0}}$&$3.6\times10^{-3\phantom{0}}$\\
Dai05 FT      &$3.4\times10^{-5\phantom{0}}$&$1.1\times10^{-6\phantom{0}}$&$3.1\times10^{-8\phantom{0}}$&---&---\\
Zoom FFT      &$3.4\times10^{-5\phantom{0}}$&$1.1\times10^{-6\phantom{0}}$&$2.5\times10^{-8\phantom{0}}$&$6.1\times10^{-5\phantom{0}}$&$4.5\times10^{-6\phantom{0}}$\\
\mr
Naïve ACF     &$3.3\times10^{-4\phantom{0}}$&$1.3\times10^{-4\phantom{0}}$&$1.2\times10^{-4\phantom{0}}$&$1.0\times10^{-2\phantom{0}}$&$4.6\times10^{-3\phantom{0}}$\\
Multi-peak ACF&$2.5\times10^{-4\phantom{0}}$&$4.2\times10^{-6\phantom{0}}$&$6.1\times10^{-8\phantom{0}}$&$6.3\times10^{-5\phantom{0}}$&$5.3\times10^{-6\phantom{0}}$\\
\mr
Gravity centre&$2.6\times10^{-5\phantom{0}}$&$6.6\times10^{-7\phantom{0}}$&$1.6\times10^{-8\phantom{0}}$&$8.9\times10^{-5\phantom{0}}$&$1.1\times10^{-5\phantom{0}}$\\
Zero crossing &$2.8\times10^{-16}$&$7.0\times10^{-16}$&$2.1\times10^{-15}$&---&---\\
Model fitting &$9.9\times10^{-17}$&$7.8\times10^{-17}$&$8.1\times10^{-17}$&$6.2\times10^{-6\phantom{0}}$&$3.2\times10^{-6\phantom{0}}$\\
\br
\end{tabular}
\end{indented}
\end{table}

\section{Good and bad practices}
\label{sec:good-bad-ugly}

About 15 methods were implemented (for 1D and 2D) and were run hundreds of thousands of times on a variety of data, ranging from one period to thousands, and with different simulated artefacts. This enabled us to draw more general conclusions and remark on the merits and pitfalls in comparison to what would be achievable from the evaluation of a small set of measured gratings.

\subsection{Evaluation of a sequence of points}
\label{sec:sequence}

When one has a sequence of key points on the profile $x_1,x_2,x_3,\dots,x_n$ (for instance zero crossings) it is tempting to compute the distances $x_2-x_1,x_3-x_2,\dots,x_n-x_{n-1}$, and average them. This would be counterproductive because only the first and last positions contribute to the average (as already noted in~\cite{Dai05}):
\begin{equation}
\hat T = \frac1{n-1} \sum_{i=1}^{n-1} (x_{i+1} - x_i) = \frac{x_n-x_1}{n-1}\;.
\end{equation}
Instead we have to fit the sequence with a straight line $x_i=iT+c$. A least squares fit gives the best unbiased linear estimate of $T$ for homoscedastic $x_n$. Nevertheless, it still gives much more weight to points close to the edges. The effective weight is proportional to the distance from the centre. Weighted fit should be considered if data close to the edges can have larger errors. This is in an interesting contrast with frequency domain methods. Although DFT itself acts uniformly, windowing suppresses data close to the edges, giving larger effective weight to data in the centre.

\subsection{Resampling}
\label{sec:resampling}

It has been suggested to interpolate the data to $K$ times larger number of points, with $K$ possibly being as high as 20~\cite{Dai05}. The results presented in this paper do not confirm this. If the evaluation method is based on data fitting (ZC and model fitting) then adding data that is  100 \% correlated with existing data cannot improve the accuracy. There is no reason to think it would improve GC either. The integrals are computed using the trapezoidal rule which already corresponds to linear interpolation. Then it is necessary to locate intersections with the threshold using interpolation~\cite{Ortlepp2021} to compute the integrals directly using non-interpolated data values. Higher order quadrature rules and higher order interpolation when finding the intersection do not bring increased precision because the trapezoidal rule approximation errors are already small compared to other error sources. On the contrary, higher order rules are prone to noise amplification. So no direct domain method benefits from resampling.

Resampling was also suggested for frequency-domain methods~\cite{Dai05}. This technique is sometimes called enhanced DFT (as opposed to refined). It is a waste of computational resources and  not recommended. As both interpolation and DFT are linear operations, the final result of the procedure is a summation over the original data $z_n$ as in eq. {\eref{dft}} but with modified weights. If the signal is sampled densely (satisfying the conditions of Nyquist--Shannon sampling theorem~\cite{Shannon49,Whittaker35}) then a perfect interpolation exists, the Whittaker--Shannon formula~\cite{Shannon49,Whittaker35}. If we simultaneously consider $K\to\infty$, \emph{i.e.}\ a very fine subdivision, equivalent to analytical evaluation of integrals, the final result simply reproduces the DFT. Worse interpolations and less precise quadratures will deviate from it (generally $Z_\nu$ would be multiplied by a slowly varying function of $\nu$), but that does not constitute an improvement. On the other hand, if the signal is undersampled then interpolation generally is not justified. It certainly cannot magically correct aliasing --- we have to measure with a shorter sampling step.

Only ACF-based methods can benefit from resampling --- moderately. They are direct space methods but average over many periods and resampling can help locating the maxima in the averaged data slightly more precisely. ACF-based methods have not been utilised in other works.

\begin{figure}
\includegraphics[width=\hsize]{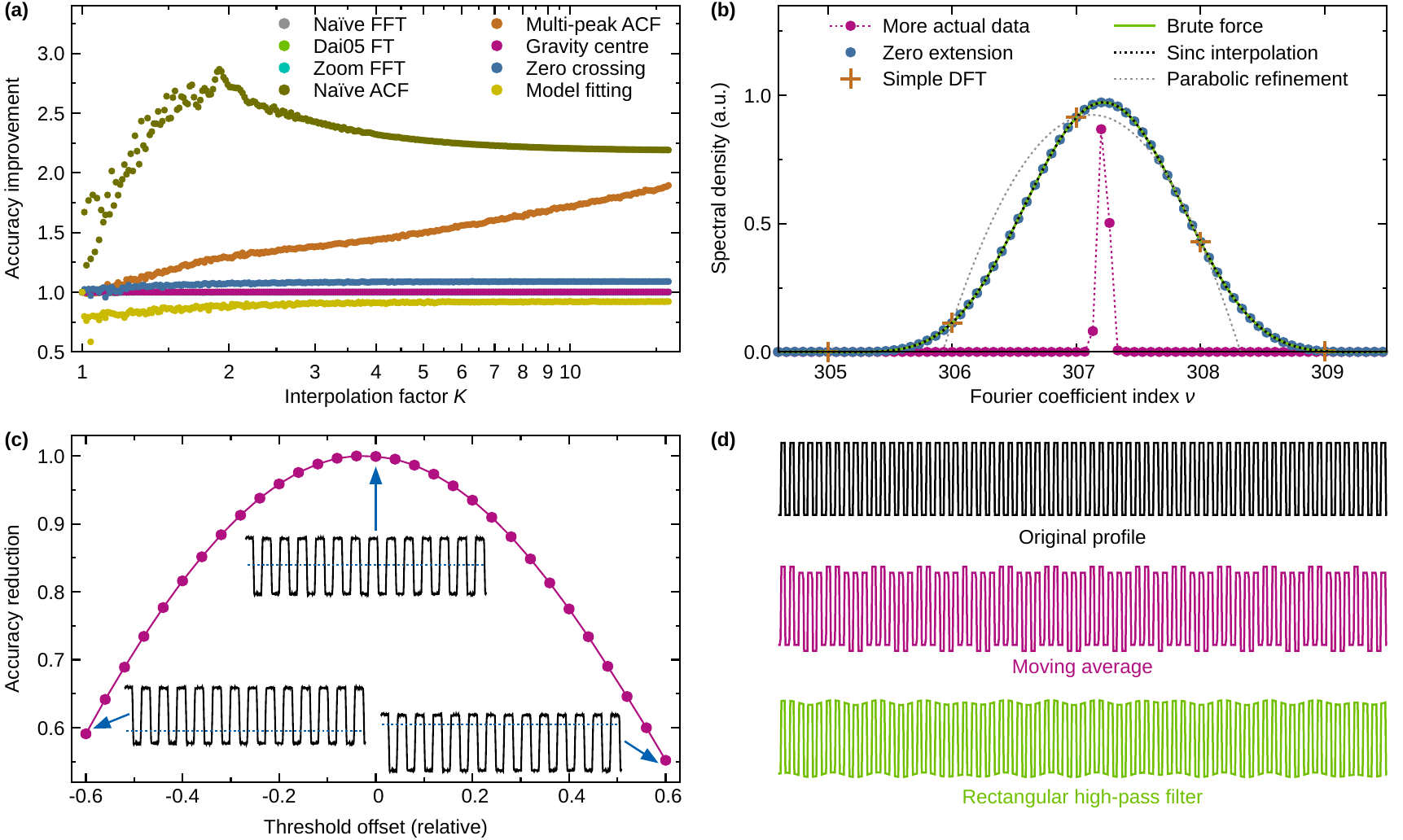}
\caption{(a) Accuracy improvement achieved by resampling the data to $K$ times larger number of points (most curves are indistinguishable from 1); (b) Fourier transform refinement; (c) reduction of accuracy in GC caused by moving the threshold away from the mid-height; (d) aliasing effects in waviness removal by linear filtering.}
\label{fig:method-remarks}
\end{figure}

The theoretical conclusions are demonstrated numerically for 1D data in \fref{fig:method-remarks}a ($N=4800$, 50 samples per period, combined noise). We calculated the improvement achieved by resampling $K$ times for all the implemented methods. Only a few curves actually differ visibly from improvement factor of 1, \emph{i.e.}\ no improvement at all. Only the two ACF methods and data fitting differ systematically. Naïve ACF shows the largest improvement, but of course from a poor base accuracy. Data fitting seems to be affected negatively. In the end, multi-peak ACF is the only method for which we might suggest resampling if greatly increasing computation time for a moderate improvement is an acceptable trade-off.

\subsection{Refined Fourier transform}
\label{sec:fft-refine}

Fourier transform refinement is the evaluation of DFT expression \eref{dft} for non-integer $\nu$, as already briefly introduced in \sref{sec:methods-1d}. It enables a more precise location of peaks in the power spectrum and thus more precise measurement of $T$~\cite{Jorgensen1998,Dai05,Ortlepp2021}. In a certain sense refinement gives the correct interpolation of $Z_\nu$ to non-integer indices $\nu$. It has been presented as substantially different from DFT~\cite{Dai05}, it has been implemented using brute force computation~\cite{Dai05,Jorgensen1998} and it was even suggested that it cannot be done by interpolating in the frequency domain~\cite{Ortlepp2021}. Some demystification is, therefore, in order.

We must start by stating the obvious: DFT is invertible. Any computation with the data $z_n$ can be reformulated with the Fourier coefficients $Z_\nu$ --- the question is only whether it is practical. DFT of almost perfectly periodic data, such as a grating profile, concentrates information into the neighbourhood of peaks in $|Z_\nu|^2$ as all other Fourier coefficients are very small, and often discardable. Therefore, Fourier coefficients from the neighbourhood are the only information that can enter the refinement, which is the definition of interpolation. A concrete illustration will be shown below, but first we have to describe how the refinement is computed using FFT.

Even if the Goertzel algorithm is used~\cite{Goertzel1958,Ortlepp2021} when Fourier coefficients are computed one by one, each still costs $O(N)$ operations. The now standard FFT-based refinement method is Zoom-FFT, a specialisation of the Chirp $z$-transform (CZT)~\cite{Shilling1970,Rabiner1969,Rabiner1975}, based on Bluestein's algorithm~\cite{Bluestein1970}. The algorithm computes $Z_\beta=\sum_n z_n s_\beta^{-n}$ for a geometric sequence of complex numbers $s_\beta$, $\beta=0,1,2,\dots,M$ (usually $s$ is denoted $z$, in line with the name $z$-transform, but it clashes with our $z$ coordinate). It does so by expressing the transform as a convolution and computing the convolution efficiently by FFT, utilising the convolution theorem. In total three FFTs of size $O(M+N)$ are needed to compute the $M$ values of $Z_\beta$. If we chose $s_\beta=\exp(-2\pi\rmi\beta/M)$, \emph{i.e.}\ uniformly covering the unit circle, we would recover DFT. But other choices are possible. In particular, we can cover only a small segment on the unit circle. This special case is called Zoom FFT because it zooms into a small interval of frequencies (although some other refinements methods are also called Zoom FFT).

Even though it may seem counter-intuitive, generally it is not recommend to locate the peak maximum using a smart search method if it comes with costly evaluation of each spectral density value. It can be efficient if the search is guaranteed to converge in $O(\log N)$ steps. However, one $N/2$ times refinement around the coarse peak and a simple scan for the maximum is much more straightforward. A parabolic refinement of the maximum can be added as a final step. Zoom FFT takes only a couple of extra FFTs and is commonly available as a function in numerical software and libraries. In higher dimensions it is more efficient to zoom twice by $\sqrt{N}$ instead of once by $N$, but in 1D there is no benefit.

Now we can look at refinement from a different angle. The transform can be refined uniformly $K$ times by computing a $K$ times larger DFT. The refined frequencies can be written $\nu+\Delta/K$, where $\Delta=0,1,2,\dots,K-1$. The refined coefficients $\bar Z$ indexed by integer $\bar\nu=K\nu+\Delta$ are then expressed
\begin{equation}
\bar Z_{\bar\nu}
= \sum_{n=0}^{N-1} z_n\exp\left(-2\pi\rmi\frac{n(\nu+\Delta/K)}{N}\right)
= \sum_{n=0}^{KN-1} \bar z_n\exp\left(-2\pi\rmi\frac{n\bar\nu}{KN}\right)\;,
\end{equation}
where $\bar z_n=z_n$ for $n<N$ and $\bar z_n=0$ for all larger $n$. In other words, the refinement is equivalently obtained by extending $z_n$ with zeros (after windowing) and computing a standard DFT. Although it would be impractical for large $K$, this formulation opens the way to interpolation in the frequency domain.

    If $K>2$ (with $K$ not necessarily an integer) the extended data satisfy the conditions of Nyquist--Shannon sampling theorem~\cite{Shannon49,Whittaker35}, except with swapped roles of direct and frequency domains. In its usual form, the theorem describes the exact reconstruction of a continuous band-limited signal using a finite number of discrete samples. However, here we have a finite-support signal in the direct domain and would like to obtain a continuous frequency spectrum from the $N$ discrete Fourier coefficients. That would be, in some sense, the correct interpolation. The interpolation is simple and given by the Whittaker--Shannon interpolation formula~\cite{Shannon49,Whittaker35}: place a \emph{sinc} function at each of the discrete frequencies, multiply it with the Fourier coefficient for this frequency and sum. Importantly, around a peak it can be realised locally with a good precision. Even though \emph{sinc} has infinite support, Fourier coefficients far from the peak contribute very little and can be disregarded. Furthermore, a spline interpolation of a reasonably high degree would probably work equally well as true \emph{sinc} interpolation~\cite{Thevenaz00}. We must emphasise that the bandwidth must be smaller than half of sampling rate. The equivalent condition here requires over half of the data to be a block of zeros. The entire construction works thanks to DFT refinement being equivalent to extending data with zeros. This is illustrated in \fref{fig:method-remarks}b for a simulated 1D grating. The large crosses show the spectral density $|Z_\nu|^2$ around the main peak (corresponding to grating pitch) computed using plain FFT. Using a brute force DFT \eref{dft} with non-integer frequencies $\nu$ we obtain the refined spectral density, which is matched perfectly by the curve obtained using \emph{sinc} interpolation. Of course, neither curve matches spectral density obtained by measuring more data. More data means more information and thus a narrower peak. The spectral density corresponding to measuring a 15 times longer profile is also shown for comparison.

A common method for refinement of the position of an extremum is to assume a parabolic shape, interpolate the three points around the coarse maximum with a parabola and take the maximum of the parabola. This, indeed, does not work as previously demonstrated~\cite{Ortlepp2021}. Peaks in DFT spectral density are frequently too narrow and cannot be interpolated by a parabola. As shown in \fref{fig:method-remarks}b, the parabolic refinement can then move the estimated maximum location in the right direction, but far from the correct position. However, after zooming sufficiently into the peak using Zoom FFT, parabolic refinement can be used.

We actually implemented the \emph{sinc} refinement in addition to the eight methods listed in \sref{sec:methods-1d}. We do not propose it as a practical method. Still, it behaves almost identically to Dai05 FT and Zoom FFT refinements and requires only a single FFT (of size $2N$) and then a bounded computation, making it one of the fastest. It is \emph{not} included in the numerical results since a third indistinguishable FT refinement would just clutter the plots.

\subsection{Zero line selection}
\label{sec:zero-line}

Zero line selection in GC and ZC attracted considerable attention in existing works. It appears as a tunable parameter~\cite{Dai05} or is located in a sophisticated manner~\cite{Ortlepp2021}. When only the upper halves are utilised it was suggested the zero line should be below the profile mid-height as larger integrated areas reduce the relative errors slightly~\cite{Huang2006}. Yet the symmetrical GC should be rather insensitive to the zero line height. The accuracy is symmetrical with respect to the optimum which occurs around the mid-height (the exact optimum is slightly different for each grating profile) and its first derivative by zero line height is zero at the optimum. Hence, as long as the zero line is approximately correct the accuracy should stay basically the same. This conclusion can already be made from~\cite{Huang2006} and is confirmed by numerical results illustrated in \fref{fig:method-remarks}c. We see that the zero line has to be moved quite far from the optimum for a substantial reduction of accuracy. It should be kept roughly around mid-height, but no great accuracy is necessary.

A similar path of reasoning can be followed for ZC. Nevertheless, we found it more sensitive to the zero line level. In addition, if the zero line is far from mid-height ZC, choosing a good segment around the crossing to fit can become more involved. The zero line was found similarly to Ortlepp \emph{et al.}~\cite{Ortlepp2021}, \emph{i.e.}\ by locating the two main peaks in height distribution and choosing the midpoint. However, this was mainly for the sake of simplicity in ZC implementation. Subtraction of the mean value was sufficient for the  other methods.

\subsection{Background subtraction}
\label{sec:background}

Subtraction of a slowly varying background on the substrate (waviness) from oscillatory data is not trivial.  A moving average~\cite{Dai05} can create a wavy pattern because the number of samples averaged is an integer, but the period is not. Sometimes the upper part of the profile contributes more to the average, sometimes the lower part, and this varies along the profile. The same conclusion can be made from analysis in the frequency domain~\cite{Ortlepp2021}. One instance of the effect is illustrated in \fref{fig:method-remarks}d. It can clearly distort an already perfectly levelled profile.

A Butterworth filter~\cite{Butterworth30}, which approximates a rectangular frequency-domain filter, was suggested as a replacement because it has monotonous frequency response~\cite{Ortlepp2021}. Unfortunately, it is not a good choice either because of its poor response to edges, where it exhibits a considerable overshoot and ringing. Furthermore, if the data processing is off-line and probably involves FFT anyway, there is no reason to limit it to filters originating in classical signal processing such as Butterworth. Filtering can be done in the frequency domain by modifying the Fourier coefficients,  even using a perfect rectangular high-pass filter, for instance. However, it would not entirely solve the poor edge response. A possible result of rectangular filter processing is also illustrated in \fref{fig:method-remarks}d.

Other linear filters, such as Gaussian or Bessel, have  abetter response. However, any linear filter is just multiplication by some function in the frequency domain and involves trade-offs between not disturbing the profile shape and removing waviness on sufficiently short length scale. In our opinion non-linear filtering may be necessary. We used a highly non-linear envelope method to remove the background. Its key feature is that it preserves an ideal rectangular wave exactly. However, selection of the optimum filter requires further investigation. 

\subsection{The lock-in method}
\label{sec:lock-in}

A lock-in method was also suggested for grating evaluation \cite{Ortlepp2021}. We did not implement and do not recommend it because it is basically a worse version of the refined FT. It proceeds as follows
\begin{enumerate}
\item Multiply the measured data by the model response --- a sine or rectangular wave is used.
\item Compute the average value. This is described in a somewhat complicated manner as `low-pass filtering', but the end result is the mean value of the multiplied data.
\item Find the maximum of this average over a domain of model response parameters, period $T$ and phase $\varphi$.
\end{enumerate}
For a sine wave the first two steps are equivalent to the computation of a Fourier coefficient. The quantity to maximise is
\begin{equation}
\sum_{n=0}^{N-1} z_n\cos\left(2\pi\frac{hn}T + \varphi\right) = \Re\left( \rme^{\rmi\varphi} Z^*_\nu\right)\;,
\end{equation}
where $\nu=Nh/T$, $\Re$ denotes the real part and $*$ complex conjugation. It attains the maximum when the absolute value of the Fourier coefficient $Z_\nu$ is maximal and $\varphi$ is equal to its phase. Therefore, the maximum coincides with the maximum of $|Z_\nu|^2$ and the method, if correctly implemented, must give the same answer as any refined FT (provided the same windowing is applied). It is, however, formulated as a multivariate optimisation problem, similar to model fitting. The analysis is a bit more complicated for rectangular waves because they contain also higher harmonics. Here we search for the combined maximum of multiple harmonics. This could in principle increase precision similarly to multi-peak ACF, even though the opposite is observed in~\cite{Ortlepp2021}. In any case such analysis, if required, would be better done in the frequency domain using a refined FT.

\subsection{Robustness and speed}
\label{sec:robustness-and-speed}

One practical concern is evaluation speed, in particular in a high-throughput context. Execution times reported in \cite{Dai05} may seem worrying, even considering the advances in computer performance. For our implementations in C using FFTW~\cite{Frigo05} and Gwyddion~\cite{Necas12} libraries we can fortunately report much more encouraging data, summarised in \tref{tab:speed}. The straightforward direct-domain methods, GC and ZC, can evaluate 50\,000 samples long profiles in a fraction of a millisecond. In 2D the two naïve methods were fastest, but they are not sufficiently accurate. The next fastest was again GC. Model fitting was the slowest and could take over 10\,s for a $1000\times1000$ image. The Gwyddion fitter has a large overhead so the fitting execution time may not be entirely representative. Still, the number of arithmetic operations per data point is invariably high in non-linear least-squares fitting.

\begin{table}
\caption{\label{tab:speed}Typical execution time of various methods, single-threaded on a standard PC. Naïve FFT and ACF are grouped under `Naïve'; more precise FFT and ACF methods under `ACF \& FFT'.}
\begin{indented}
\item[]\begin{tabular}{@{}lccccc}
\br
Data&GC \& ZC&Naïve&FFT \& ACF&Dai05 FFT&Model fitting\\
\mr
profile, 50k points&0.2 ms&2 ms&5 ms&30 ms&300 ms\\
image $1000\times1000$&100 ms&40 ms&0.5 s&---&10 s\\
\br
\end{tabular}
\end{indented}
\end{table}

Perhaps an even more important property is method robustness, \emph{i.e.}\ ability to behave correctly for a wide range of input data.
In the simulations all methods were run hundreds of thousands times on generated data without human intervention, demanding perhaps a bit more robust implementations than is typical. With FFT and ACF based computations this was easy to achieve since they work with transformed (`summary') data and are insensitive to local defects. If there is a peak where the algorithm is looking for it, it is found and measured correctly. This makes them very reliable.

Model fitting is decidedly less robust. Its known Achilles' heel is initial parameter estimation and the existence of multiple local minima of the sum of squared residuals. Naïve FFT was used for initial estimation of $T$ and a scalar product similar to the lock-in method (\sref{sec:lock-in}) for phase estimation. Such an estimate is still insufficient for long gratings --- when the estimated $T$ is 1\,\% off the model gets completely out of phase after 50 periods. Therefore, only several periods were initially fitted and the fitted segment was increased in a geometric progression until the entire data were covered. Multiple local minima did not seem a major problem with long periodic data, such as gratings. Although fitting can definitely fail when the initial estimate is not sufficiently close, we observed more or less only two possible outcomes. It either succeeded and gave an accurate $T$ value --- or failed rather obviously. We should also mention that model fitting does not work without a reasonable parametric description of the data. Gratings are relatively simple to describe, but other periodic structures may be more complicated. Evaluation of a different type of sample may require the construction of a new model.

Bootstrapping was necessary also in other methods. All tested background removal procedures require at least an approximate \emph{a priori} knowledge of the period. Furthermore, any refined FT method starts from a coarse estimate. Naïve FFT is reliable and serves well for this purpose, unless a very small number of periods is measured. In such case the profile may have to be zero-extended (after windowing) to two or three times the size to increase the frequency resolution.

GC and ZC require the most care to work reliably. Particles and other local defects can lead to incorrect bar gravity centres~\cite{Dai05}. When $N/P$ is high, noisy data can cross the threshold more than once, creating very short segments which need to be filtered out~\cite{Ortlepp2021}. Our GC implementation used a two-stage filtering. Most incorrect segments come from multiple threshold crossings and are too short, whereas too long are rare. Therefore, the first stage computed the 90 th percentile of segment lengths and segments more than $10\times$ shorter were discarded (and segments tat were too long). The second stage found the median inter-centre distance and only kept centres whose distance to the closest good centre was close to an integer, avoiding possible problems illustrated in figure 13 in \cite{Dai05}. The initial set of good centres was identified as three consecutive centres with distances close to the median. ZC points were processed in the same manner, just without the first stage because an entire region around a crossing is fitted and so each crossing only gives one value of $x_0$. This approach made the procedures self-contained. If a FFT-based estimate of $T$ is available, the filtering can be simplified. A modification of random sample consensus (RANSAC)~\cite{Fischler81} may also be suitable. In both the GC and the ZC cases, the positive and negative features were separately fitted and their average take in order to reduce the sensitivity of the methods to the choice of threshold.

GC and ZC employ various thresholds, interval/point filtering parameters, fitting ranges and other similar algorithm tunables. All could be made user-controllable, as was suggested for the thresholds~\cite{Dai05,Huang2006}. One has to resist the temptation do so. In addition to the usual problems that follow~\cite{Necas17} we noticed one specific to highly accurate measurements. It is not difficult to make the evaluation accuracy worse by an order of magnitude or two by a poor parameter choice (or a subtle implementation error). What may be difficult for the user, is spotting that the relative error jumped from $3\times10^{-7}$ to $3\times10^{-5}$, even in a situation when the former is negligible and the latter a major error source. Extensive numerical verification can ensure a method achieves the accuracy it should when used for samples within defined parameters. With half a dozen tunables the user can tweak freely this becomes impossible. From this point of view refined FT based methods are preferable.

\section{Conclusion}

\begin{table}
\caption{\label{tab:pros-and-cons}{Summary of advantages and disadvantages of various methods.}}
\begin{indented}
\item[]\begin{tabular}{@{}lll}
\br
Method&Advantages&Disadvantages\\
\mr
Manual&easy to understand&poor accuracy upper limit\\
      &&poor representativeness\\
      &&user-dependent\\
      &&laborious\\
\mr
Naïve FFT&insensitive to local defects&poor accuracy\\
         &insensitive to background&no error estimate\\
         &insensitive to odd shapes&\\
         &relatively fast&\\
\mr
Refined FT&good accuracy&no error estimate\\
          &insensitive to local defects&\\
          &insensitive to background&\\
          &insensitive to odd shapes&\\
\mr
Naïve ACF&insensitive to local defects&poor accuracy\\
         &insensitive to background&no error estimate\\
         &relatively fast&sensitive to shape asymmetry\\   
\mr
Multi-peak ACF&good accuracy&sensitive to shape asymmetry\\
              &insensitive to local defects&relatively low accuracy upper limit\\
              &insensitive to background&\\
              &provides error estimate&\\
              &relatively fast&\\
\mr
Gravity centres&good accuracy&difficult to make robust\\
               &provides error estimate&depends on background subtraction\\
               &fast&\\
\mr
Zero crossings&good accuracy&difficult to make robust\\
              &provides error estimate&depends on background subtraction\\
              &high accuracy upper limit&only one-dimensional\\
              &fast&\\
\mr
Model fitting&good accuracy&requires a model\\
             &provides error estimate&depends on parameter estimation\\
             &high accuracy upper limit&slow\\
             &possibly multiple parameters&\\
\br
\end{tabular}
\end{indented}
\end{table}

In this paper we report on the use of both direct space and Fourier space based methods for evaluating the periodic structure parameters. In addition to gravity centres and refined FT methods described in the ISO standard, this includes zero crossing, multi-peak ACF and model fitting. We concluded that  these five evaluation methods can be recommended, if implemented properly. Overall they all behave similarly, although some differences in sensitivity to various artefacts exist; see table \ref{tab:pros-and-cons}.  Importantly, their accuracy scales super-linearly with the number of periods $P$, typically with $P^{3/2}$ for profiles and $P^2$ for images. As a side effect of this analysis, more can be said about the overall benefits and drawbacks of different methods, as also shown in table \ref{tab:pros-and-cons}.

There are also more general conclusions and recommendations that can be used when designing the experiment and processing the measured data, these are explained in the previous text and are summarized here:
\begin{itemize}
\item Since all the good methods are sub-pixel, pixel size/sampling step is not the limiting factor for accuracy.
\item Although the ISO standard recommends measuring more than 5 or 7 samples per period, ideally at least 20 points should be measured. If at least 20 pixels per period can be measured, measure as many periods in the profile as possible.
\item Resampling measured data offers no benefits.
\item Accuracy scaling with the $3/2$-th power of number of periods is quite difficult to beat using alternative measurement strategies. The accuracy scales only linearly with decreasing sampling step and only with the square root of the number of repetitions.
\item The GC and ZC methods readily provide an estimate of the statistical error.
\item If these guidelines are followed, the uncertainty contribution from the numerical procedure used is insignificant compared to the other uncertainty components related to  measurement (at least in the case of a standard SPM). 
\end{itemize}

As a simple example of applying the above recommendations, for most typical gratings used in a commercial AFM calibration, which are 1--\SI{5}\um{} pitch, use full range of the microscope (typically \SI{100}\um) as long as the positioning errors of the scanning stage are not significant at the periphery. These settings should provide 20--100 periods. Keeping at least 20 pixels per period then means collecting about 2000 pixels per line, which is achievable by standard SPMs.

It is hoped that the results presented in this paper will give users confidence in evaluating periodic structures and calibrating SPMs.


\subsection*{Acknowledgements}

This work was supported by Technology Agency of the Czech Republic project No. TN01000008 and by the project 20IND08 MetExSPM that received funding from the EMPIR programme co-financed by the Participating States and from the European Union’s Horizon 2020 research and innovation programme. Part of this work was funded by the National Measurement System Programme from the Department of Business Energy and Industrial Strategy, UK.

\subsection*{References}
\bibliographystyle{iopart-num}
\bibliography{article}

\end{document}